\documentclass[11pt,a4paper]{article}
\pdfoutput=1

\usepackage{jheppub}
\usepackage{color}
\usepackage[dvipsnames]{xcolor}
\usepackage{graphicx}
\usepackage{wrapfig}
\usepackage{verbatim}
\usepackage{amsmath}
\usepackage{amssymb}
\usepackage{url}
\usepackage{bbold}
\usepackage{xspace}
\usepackage{slashed}
\usepackage{multirow}
\usepackage{threeparttable}
\usepackage{paralist}
\usepackage{subcaption}
\usepackage{floatrow}
\usepackage{multirow}
\usepackage{soul}
\usepackage[export]{adjustbox}
\usepackage[normalem]{ulem}
\usepackage{siunitx}

\usepackage{tikz}
\usetikzlibrary{trees}
\usetikzlibrary{decorations.pathmorphing}
\usetikzlibrary{decorations.markings}
\usetikzlibrary{arrows}

\newcommand{\MeV}{\,\mathrm{MeV}}
\newcommand{\GeV}{\,\mathrm{GeV}}
\newcommand{\TeV}{\,\mathrm{TeV}}

\newcommand{\fm}{\,\mathrm{fm}}

\newcommand{\vev}[1]{\langle #1 \rangle}

\newcommand{\beq}{\begin{equation}}
\newcommand{\eeq}{\end{equation}}
\newcommand{\bea}{\begin{eqnarray}}
\newcommand{\eea}{\end{eqnarray}}

\DeclareRobustCommand{\Sec}[1]{Sec.~\ref{#1}}
\DeclareRobustCommand{\Secs}[2]{Secs.~\ref{#1} and \ref{#2}}
\DeclareRobustCommand{\App}[1]{App.~\ref{#1}}

\DeclareRobustCommand{\Fig}[1]{Fig.~\ref{#1}}
\DeclareRobustCommand{\Figs}[2]{Figs.~\ref{#1} and \ref{#2}}
\DeclareRobustCommand{\Eq}[1]{Eq.~(\ref{#1})}

\DeclareRobustCommand{\Refs}[1]{Refs.~\cite{#1}}

\newcommand{\km}{\,\mathrm{km}}
\newcommand{\ie}{i.e.~}
\newcommand{\eg}{e.g.~}
\newcommand{\Tideal}{T_{\text{\tiny ideal}}}
\newcommand{\Tgrad}{T_{\text{\tiny grad}}}
\newcommand{\mpl}[1]{M^{#1}_\text{\tiny P}}
\newcommand{\mphi}[1]{M^{#1}_\phi}

\newcommand{\ngs}{\text{\tiny NGS}}
\newcommand{\dm}{\delta m_*}
\newcommand{\zud}{z_{ud}}

\begin{document}

\hfill IFT-UAM/CSIC-23-96

\hfill TUM-HEP-1468/23

\vspace{-1.5cm}

\title{Heavy neutron stars from light scalars}

\author[a]{Reuven Balkin,}
\author[b]{Javi Serra,}
\author[c]{Konstantin Springmann,}
\author[d]{Stefan Stelzl,}
\author[c]{Andreas Weiler}

\affiliation[a]{Physics Department, Technion -- Israel Institute of Technology, Haifa 3200003, Israel}
\affiliation[b]{Instituto de F\'isica Te\'orica UAM/CSIC, Madrid 28049, Spain}
\affiliation[c]{Physik-Department, Technische Universit\"at M\"unchen, 85748 Garching, Germany}
\affiliation[d]{Institute of Physics, Theoretical Particle Physics Laboratory, \'Ecole Polytechnique F\'ed\'erale de Lausanne, CH-1015 Lausanne, Switzerland}


\date{\today}

\abstract{
We study how light scalar fields can change the stellar landscape by triggering a new phase of nuclear matter.
Scalars coupled to nucleons can develop a non-trivial expectation value at finite baryon density. 
This sourcing of a scalar reduces the nucleon mass and provides an additional energy density and pressure source. 
Under generic conditions, a new ground state of nuclear matter emerges, with striking implications for the configuration of stellar remnants. 
Notably, neutron stars in the new ground state can be significantly heavier than QCD equations of state currently predict. 
We also find hybrid stellar compositions and stable self-bound objects with sizes as small as the Compton wavelength of the scalar. 
We discuss several specific realizations of this scenario: the QCD axion and lighter generalizations thereof and linearly or quadratically coupled scalar fields effectively equivalent to a class of scalar-tensor modification of gravity. 
Lastly, we explore phenomenological signatures relevant to electromagnetic and gravitational wave observations of neutron stars, such as atypical compactness and instability gaps in radii.}

\preprint{}
\maketitle


\section{Introduction}
\label{sec:intro}

The extremely rich interplay between light scalar fields beyond the Standard Model (BSM) and compact stellar objects has recently generated increasing interest due to advances on both theoretical and experimental fronts. 
On the theory side, light scalars are a common prediction of BSM theories that address some of the standing problems of the SM. 
Some well-known examples are the QCD axion as a solution to the strong CP problem \cite{Weinberg:1977ma,Wilczek:1977pj,Peccei:1977hh,Vafa:1984xg}, landscape scalars from cosmological solutions to the electroweak hierarchy problem as in \cite{Graham:2015cka,Arvanitaki:2016xds,Arkani-Hamed:2020yna,TitoDAgnolo:2021pjo}, and scalar-tensor theories classified as modifications of gravity in the infrared \cite{Fierz:1956zz,Jordan:1959eg,Brans:1961sx,Horndeski:1974wa,Damour:1992we,Damour:1993hw,Damour:1994zq,Khoury:2003aq,Nicolis:2008in,Hinterbichler:2010es,Zumalacarregui:2013pma,Gleyzes:2014dya,Langlois:2015cwa}, see also \cite{Langlois:2018dxi,Sakstein:2018fwz,Brax:2021wcv,Doneva:2022ewd} for recent reviews.

Especially when the scalars are coupled weakly to matter, compact objects in general and neutron stars (NSs) in particular provide a great environment to study their dynamics, owing to the large number (density) of SM particles. 
This has been widely recognized in the past, with many instances of  interesting physical processes that take place as a result: new sources of energy loss that affect cooling rates, supernovae dynamics, electromagnetic emissivity and overall evolution \cite{Iwamoto:1984ir,Turner:1987by,Raffelt:1987yt,Raffelt:1990yz,Raffelt:2006cw,Payez:2014xsa,Sedrakian:2015krq,Beznogov:2018fda,Chang:2018rso,Carenza:2019pxu,Buschmann:2019pfp,Buschmann:2021juv}, superradiance in stars \cite{Arvanitaki:2009fg,Cardoso:2015zqa,Endlich:2016jgc,Day:2019bbh,Kaplan:2019ako,Chadha-Day:2022inf}, (un-)screening or emergence of fifth-forces via spontaneous scalarization \cite{Damour:1993hw,Damour:1994zq,Khoury:2003aq,Hinterbichler:2010es,Brax:2012gr,Ramazanoglu:2016kul,Hook:2017psm,Staykov:2018hhc,Blinov:2018vgc,Doneva:2022ewd}, modifications of the merger of NSs and their electromagnetic and gravitational-wave (GWs) signatures \cite{Barausse:2012da,Shibata:2013pra,Sagunski:2017nzb,Huang:2018pbu,Harris:2020qim,Zhang:2021mks,Dev:2021kje}, 
alteration of pulsar dynamics \cite{Taylor:1982zz,Damour:1991rq,Damour:1996ke,Freire:2012mg,Will:2014kxa,Hook:2017psm,Archibald:2018oxs,Kaplan:2019ako} and even induced vacuum decay \cite{Blinov:2018vgc,Hook:2019pbh,Balkin:2021zfd,Balkin:2021wea}.

Another exciting yet relatively unexplored aspect, which is the focus of this paper, concerns the structure and composition of stellar remnants when a scalar field develops a non-trivial profile in medium. 
Such a scalarization takes place for dense and large enough objects. This translates into an upper bound on the scale $\mphi{}$ which, in analogy to $\mpl{}$, controls the strength of the scalar interaction with matter, along with an upper bound on its mass $m_\phi$. 
These conditions in turn imply that the scalar is not significantly sourced, or not sourced at all, by dilute and small systems, nor arbitrarily long-ranged, such that conventional constraints from fifth-force searches and e.g.~pulsar timing measurements can be bypassed, even for $\mphi{} \ll \mpl{}$.
Particularly in the context of modified theories of gravity, it has been recognized that scalarized NSs exhibit macroscopic properties, like mass, radius, moment of inertia, etc., different from those predicted in GR with a given equation of state (EOS), see e.g.~\cite{Doneva:2022ewd} for a review. 
However, it has not been fully appreciated until now that, under generic conditions, large deviations in the configuration of compact stars are due to the appearance of a new ground state (NGS) of nuclear matter.
\footnote{This is with the exception of Ref.~\cite{Gao:2021fyk}, which however considered a scalar linearly coupled to matter without screening, thus generically ruled out. 
Besides, the implications of the NGS on the configuration of white dwarfs were studied by the authors in Ref.~\cite{Balkin:2022qer}, showing how current measurements of their radii can be used to exclude the parameter space of lighter deformations of the QCD axion.}
In this work, we explain how such a scalarized ground state of matter can be reached when the higher-order 
interactions of the scalar allow it to approach $\phi/\mphi{} = O(1)$ inside the star.

To make the physics as transparent as possible, we carry out our analysis in terms of the EOS of a free Fermi gas of nucleons. 
In the standard picture, Fermi degeneracy pressure stabilizes the star against gravitational collapse. 
The existence of a scalar coupled (non-derivatively) to nucleons, which can be encoded as a scalar-dependent nucleon mass, $m_*(\phi) \bar\psi \psi$, brings up two competing effects when the conditions for scalarization are met. 
On the one hand, the effective mass of the nucleon inside the scalar bubble gets reduced \cite{Ellis:1989as}, while on the other, the scalar potential $V(\phi)$ acts as additional vacuum energy contributing to the total energy density and pressure, see e.g.~\cite{Bellazzini:2015wva,Csaki:2018fls}. 
The parameter space is then generically split into two parts: if the potential dominates, a phase transition occurs at some finite density, softening the EOS. 
Stable configurations are in the form of hybrid stars, with smaller maximal masses compared to the phase where the scalar field is not sourced. 
In the rest of the parameter space, the reduction of the nucleon mass dominates and a new ground state of matter, with energy per particle smaller than for well-separated neutrons, is found. 
Stable configurations are in the form of homogeneously scalarized stars, which can have much larger masses. NSs in this case can be heavier than the maximal mass predicted by standard causal bounds \cite{1973ApJ...179..277N,Rhoades:1974fn,Kalogera:1996ci,Lattimer:2012nd}, which assume a certain low-density behavior consistent with the properties of dilute matter, an invalid assumption when the NGS is present. 
Small or dilute systems like nuclei or regular stars are in this case meta-stable and long-lived. 
Other distinctive features of NSs in the NGS are their compactness, which can be larger than in typical NSs yet below that required for a photon sphere, and their minimal rotation period. 
In addition, since the NGS leads to a branch of NSs disconnected from the standard one predicted in GR, we find instability gaps in radii and self-bound objects as small as the Compton wavelength of the scalar.

On the experimental side, the observables mentioned above are particularly exciting in the advent of recent and future multi-messenger astronomy. 
The most prominent instance of this fact is the binary NS merger GW170817 detected in GWs by LIGO and Virgo \cite{LIGOScientific:2017vwq} along with an electromagnetic counterpart (GRB 170817A) detected by Fermi gamma-ray telescope \cite{LIGOScientific:2017zic} which, among many other physics results, led to constraints on NS radii, maximal mass, and EOS, see e.g.~\cite{LIGOScientific:2018cki,Margalit:2017dij,Annala:2017llu,Shibata:2019ctb}. 
This is promising in light of the many more NS-NS as well as NS-BH (black hole) mergers that the current network of GW detectors is expected to detect \cite{KAGRA:2013rdx}. 
The current stellar remnant catalog \cite{LIGOScientific:2018mvr,LIGOScientific:2020ibl,LIGOScientific:2021djp} will also be significantly expanded by third-generation GW observatories \cite{Punturo:2010zz,Reitze:2019iox}.
Another interesting result is the merger event GW190814 \cite{LIGOScientific:2020zkf}, which measured one of the progenitors to be a stellar remnant with a mass of approximately $2.6M_{\odot}$. 
This could be the heaviest NS or the lightest BH discovered to date. While a distinction between NSs and BHs from GW data alone is non-trivial, see e.g.~\cite{Brown:2021seh,Coupechoux:2021bla}, this event poses a challenge to the expected lower mass gap between NSs and BHs \cite{Bednarek:2011gd,Oertel:2016bki,Bombaci:2016xzl,Lattimer:2021emm,Nathanail:2021tay,Bailyn:1997xt,Fryer:1999ht,Ozel:2010su}. 
While theoretical uncertainties regarding both the maximal mass of NSs as well as the minimal mass of BHs are large, the possibility of heavy scalarized NSs provides a potential avenue to fill the gap, in particular given their violation of standard maximal-mass bounds.
Furthermore, electromagnetic measurements are  essential to the study of NSs.
Observations of radio pulsar timing give information on the maximal mass of NSs \cite{NANOGrav:2019jur,Fonseca:2021wxt} and, in the future, potentially on their moment of inertia \cite{Hu:2020ubl}, leading to important constraints on NS matter. 
NICER measurements in X-rays from binary pulsars have already improved the radius determination of NSs  \cite{Miller:2019cac,Riley:2021pdl}, and significant progress in this type of observational techniques as well as in theoretical modelling is expected in the near future, see e.g.~\cite{Ozel:2016oaf,Bogdanov:2022faf}.

The paper is organized as follows. 
In \Sec{sec:model}, we start by presenting our simple free Fermi gas description and the equations governing the full coupled system of gravity plus the scalar. 
All of our results are described in this section in a model-independent fashion. 
In \Sec{subsec::micro_EOS_neg_gradient}, we discuss the limit where the scalar field gradient can be neglected, in which case it is sensible to define an EOS. 
After presenting the two qualitatively different types of EOS, we discuss the effects of a finite gradient energy in \Sec{subsec::finite_f_effects}.  
A more quantitative case study in presented in \Sec{sec::case_studies} for three types of scalar-matter couplings, namely axion-like (\Sec{sec::bounded_models}), linear (\Sec{sec::linear_coup_model}), and quadratic (\Sec{sec::quad_coupling}), while in \Sec{sec::scalar-tensor} we explicitly work out the equivalence with scalar-tensor theories of gravity. 
Our conclusions are presented in \Sec{sec:conc}. 
A detailed discussion regarding the limit of negligible gradient energy, which is used extensively throughout this work, is presented in \App{app::dim_anal_neg_grad}. 
Some useful analytical approximations for constant density objects are given in \App{app::object_class}. 
Finally, in \App{app::fGG} we present the details on a simple scalar model that realizes a large in-density reduction of the nucleon mass via a coupling to gluons.


\subsection{Decoding scalarization}
\label{subsec::chatgpt}

Before moving on to the bulk of our work, let us briefly discuss the main features of the different classes of scenarios in which a scalarized ground state of matter can be reached. 
Beyond the scale $\mphi{}$ that sets the strength of the leading coupling to matter and the mass $m_\phi$, the scalar theories under consideration can be characterized in general by two other scales, $F_\phi$ and $f_\phi$, which control, respectively, the higher-order interactions of the scalar with matter and its self-interactions,
\begin{equation}
m_*(\phi)/m = 1 - \bigg( \frac{\phi}{\mphi{}} \bigg)^n \left[ 1 + O \left(\frac{\phi}{F_\phi}\right) \right] \, , \quad V(\phi) = \frac{1}{2} m_\phi^2 \phi^2 \left[ 1 + O \left(\frac{\phi}{f_\phi}\right) \right] \, .
\label{eq:mV}
\end{equation}
where $m_*(\phi) \bar\psi \psi$ is the scalar-dependent matter mass. We focus only on linearly ($n = 1$) or quadratically ($n = 2$) coupled scalars, as these two are the most generic cases without or with a parity symmetry $\phi \to -\phi$ in the interaction with matter, respectively. 
The simplest scenario is when higher-order terms in $\phi$ can be neglected, namely when $F_\phi, f_\phi \gg \mphi{}$.
In that case scalarization takes place with $\phi \sim \mphi{}$ and $m_*(\mphi{})/m \ll 1$ is approached within the star, a limit in which the scalar field effectively reaches a constant in-medium value.
This case, which we denote as \emph{unbounded} $m_*$, is typical of scalar-tensor theories of the Damour-Esposito-Far\`ese type \cite{Damour:1993hw}, where the scalar couples to the trace of the energy-momentum tensor, equal to $m \bar \psi \psi$ in the free Fermi gas limit. 
Factoring in higher $\phi$ terms, even when $f_\phi < \mphi{}$, one can still find unbounded $m_*$ systems where $\phi \sim \mphi{}$ in medium, specially at high-enough densities where the scalar dynamics is mainly controlled by $m_*(\phi)$ (with $F_\phi \gg \mphi{}$). 
In this class of scenarios, the main effect of a finite $f_\phi$ is to set the value of the scalar potential $V(\mphi{})$. 
Finally, there exists a \emph{bounded} $m_*$ class of theories, in which non-linearities are such that $\phi \sim F_\phi$ or $\phi \sim f_\phi$ inside the star. 
The typical example is chameleon screening \cite{Khoury:2003aq} (see also \cite{Blinov:2018vgc}), where non-linear terms in the potential prevent the scalar from reaching in-medium values much beyond $\phi \sim f_\phi \ll \mphi{}$. 
In this paper, we pay more attention to scenarios where non-linearities in the scalar-matter interactions force $\phi \sim F_\phi \ll \mphi{}$ inside NSs. 
This is the case (for $n=2$) of the QCD axion, if sourced by dense matter \cite{Balkin:2020dsr}, and by its generalizations where the axion coupling to gluons is decorrelated from its mass \cite{Hook:2017psm,Hook:2018jle,DiLuzio:2021pxd}.
We will not discuss ($n=1$) scalar-tensor theories of the Damour-Polyakov type \cite{Damour:1994zq} since in that case $m_*(F_\phi) = m$.

Finally, let us note that we are leaving aside theories in which scalarization takes place not because of the interactions with matter but due to the coupling of the scalar to the curvature (\ie to $R_{\mu \nu \rho \sigma}^2$ at leading order in a derivative expansion) \cite{Yunes:2009hc,Sotiriou:2013qea,Silva:2017uqg}. 
See however \cite{Serra:2022pzl,Hong:2023zgm} for recent theoretical constraints on this type of scenarios.


\section{Scalarized free Fermi gas}
\label{sec:model}

Neutron stars are well-described at leading order by a degenerate free Fermi gas coupled to gravity.  To study the effects of a scalar field coupled to nucleons, we consider the following Lagrangian, containing a single massive fermion $\psi$ and a single real scalar $\phi$, coupled to the gravitational field $g_{\mu\nu}$,
\begin{align}
\label{eq:lagr}
\mathcal{L}_{\psi \phi} = \sqrt{-g}\left[\bar{\psi} \left(ie^{\mu}_{\,\,\,a}\gamma^a D_\mu -m_*(\phi)\right)\psi+\frac12 g^{\mu\nu} (\partial_\mu \phi)(\partial_\nu \phi) -V(\phi)\right]\,,
\end{align}
where $D_\mu = \partial_\mu - i \omega_\mu$ is the covariant derivative of a fermion field in curved space. $\psi$ and $\phi$ are coupled via the term $m_*(\phi) \bar\psi \psi$, while the self-interactions of $\phi$ are encoded in the function $V(\phi)$, see \Eq{eq:mV}. Such a coupling to nucleons naturally arises in models where the scalar field couples in the UV to the (light) quarks, or to the gluon field strength,  or simply to the trace of the energy-momentum tensor (see \Sec{sec::scalar-tensor}). For convenience, we shall henceforth work with the dimensionless field $\theta \equiv \phi/f$, where we introduce the scale $f$ as the typical scale of the scalar field. This can be conveniently identified with either $\mphi{}$, $F_\phi$, or $f_\phi$ introduced in \Eq{eq:mV}, depending on the particular realization. We further assume that at zero density (\ie in the absence of the Fermi gas), the potential $V(\theta)$ is minimized at $\theta_0$ such that
\begin{align}
\frac{\partial V}{\partial \theta}\bigg|_{\theta=\theta_0} =0\,, \;\;\; V(\theta_0)=0\,, \;\;\; \text{and} \;\;\; m(\theta_0)\equiv m\;\;\; \text{(at zero density)}\,.
\end{align}
Let us derive the static coupled equations of motion (EOMs) for the fermion, scalar and gravitational fields. Assuming radial symmetry, we use the Schwarzschild metric parametrization, 
\begin{equation}
g_{00}=e^{2\nu(r)} \,, \;\;\; g_{rr} = -\left[1-\frac{2M(r)}{ \mpl{2} r} \right]^{-1}\, \;\;\; g_{\Theta\Theta} = -r^2\,, \;\;\; g_{\varphi\varphi} = -r^2\sin^2\Theta\,,
\end{equation}
where $\mpl{} = G^{-1/2}$ and we choose the metric convention $\eta_{\mu\nu} = \text{Diag}[1,-1,-r^2,-r^2\sin^2\Theta]$. The gravitational field is sourced by an energy-momentum tensor that is the sum of two terms,
\begin{align}
T^{\mu}_{\;\;\;\nu}= (\Tideal)^{\mu}_{\;\;\;\nu}+(\Tgrad)^{\mu}_{\;\;\;\nu}\,.
\end{align}
$ \Tideal$ contains the contributions of the Fermi gas and the scalar potential $V(\theta)$, and has the form of an ideal fluid, \ie $(\Tideal)^{\mu}_{\;\;\;\nu} = \text{Diag}[\varepsilon,-p,-p,-p]$ with
\begin{align}
\label{eq::epsandptot}
\varepsilon  &= \varepsilon_{\psi}(m_*(\theta),\rho)+V(\theta) \;\;\;\;\text{and}\;\;\;\; p =p_{\psi}(m_*(\theta),\rho)-V(\theta) \,.
\end{align}
The total pressure of the system $p$ as defined above can become negative in regions where the contribution from the potential $V(\theta)$ dominates over the strictly positive pressure of the Fermi gas.
Note also that $\varepsilon_{\psi}, p_{\psi}$ and $\varepsilon, p$ separately satisfy the thermodynamic relation $d(\epsilon/\rho) = - p d(1/\rho)$ in the constant $\theta$ limit.
$(\Tgrad)^{\mu}_{\;\;\;\nu}$ is proportional to $f^2$,
\begin{align}
(\Tgrad)^{\mu}_{\;\;\;\nu}  = f^2(\partial_r \theta)^2  \left[ \frac12  \delta_{\nu}^\mu-\delta^{\mu}_{r}\delta_{\nu}^{r}\right] \left(1-\frac{2M}{r \mpl{2}}\right)\,.
\end{align}
The term in the square brackets describes the gradient energy of the field and has the form of a perfect fluid. The second term deviates from the perfect fluid behavior in the form of additional (anisotropic) pressure. Both terms are proportional to $f^2$,  therefore we expect them to become negligible when $f$ is much smaller than other scales appearing in the EOM, see \App{app::negligible_grad} for a detailed discussion. We derive three independent EOMs by minimizing the action defined by the Lagrangian $\mathcal{L} = ({\mpl{2}}/16\pi)\sqrt{-g}R+\mathcal{L}_{\psi \phi}$ to find
\begin{subequations}
\label{eq::TOV}
\begin{align}
\theta''\bigg(1\bigg.&\bigg.-\frac{2M}{r \mpl{2}}\bigg)+\frac{2}{r}\theta'\left(1-\frac{M}{r \mpl{2}}-\frac{ 2\pi r^2}{ \mpl{2} } \left(\varepsilon-p  \right)\right) = \frac{1}{f^2}\left( \frac{\partial V}{\partial \theta}+\rho_s \frac{\partial m_*(\theta)}{\partial \theta}\right),\label{eq::gen_KG}\\
\begin{split}
p'=&-\frac{M\varepsilon}{ \mpl{2} r^2}\bigg[1+\frac{p}{\varepsilon} \bigg]\left[1-\frac{2M}{r \mpl{2}}\right]^{-1}\left[1+ \frac{4\pi r^3}{M}\left(p+\frac12 f^2  {\theta'}^2\left\{1-\frac{2M}{r \mpl{2}}\right\}\right) \right]\\
 &-\theta' \left(\frac{\partial V}{\partial \theta}+\rho_s \frac{\partial m_*}{\partial\theta} \right),
\end{split}\label{eq::TOV1}\\
M' =&\,\, 4\pi r^2  \left(\varepsilon+\frac12 f^2  {\theta'}^2\left[1-\frac{2M}{r \mpl{2}}\right]\right),\label{eq::TOV2}
\end{align}
\end{subequations}
where we introduced the fermion scalar density $\rho_s(m_*(\theta),\rho) \equiv \langle \bar\psi \psi \rangle $.
\Eq{eq::gen_KG} is the generalized form of the scalar EOM. It contains the coupling to gravity, which deforms the derivatives on the LHS.
The scalar self-interactions are encoded in the first term in parenthesis on the RHS. The scalar interaction with the nucleons is given by the second term in parenthesis on the RHS.
The last two equations are the generalized Tolman-Oppenheimer-Volkoff (TOV) equations~\cite{Oppenheimer:1939ne,Tolman:1939jz}.
\Eq{eq::TOV1} dictates how the total pressure is balanced by the gravitational force and an additional new scalar force. \Eq{eq::TOV2} is associated with the enclosed mass $M(r)$, found by integrating over the energy density associated with the Fermi gas, the scalar potential, as well as a contribution from the scalar gradient.
There are in principle two additional EOMs which we do not present. The EOM for the fermion field is implicitly used in the expression for the energy and pressure of the Fermi gas.\footnote{Note that one should solve the fermion EOM in flat space, such that the microscopic properties of the fermion gas are independent of the gravitational field; one can always choose a reference frame which is flat at the characteristic scales of the Fermi gas.} 
The equation of motion for the temporal component of the metric $\nu(r)$ can be solved separately since $\nu(r)$ and its derivatives do not appear in any of the other equations. The combination of all EOMs imply by construction energy-momentum conservation, \ie $\partial_{;\mu} T^{\mu\nu} =0 $, which is equivalent to the so-called hydrostatic equilibrium condition~\cite{Weinberg:1972kfs}. 
\\ \\
The derivation so far has been independent of the properties of the Fermi gas. For concreteness, from this point on we consider the simple case of a free Fermi gas (see \eg \cite{Glendenning:1997wn})\begin{subequations}
\begin{align}
\varepsilon_{\psi}(m_*(\theta),\rho) &= 2\int^{k_F(\rho)}\frac{\mathrm{d}^3k}{(2\pi)^3} \sqrt{{\bf k}^2+m_*^{2}(\theta)}\,,
\\
p_{\psi}(m_*(\theta),\rho) &= \frac{2}{3}\int^{k_F(\rho)}\frac{\mathrm{d}^3k}{(2\pi)^3} \frac{k^2}{\sqrt{{\bf k}^2+m_*^{2}(\theta)}}\,,
\\
\rho_s(m_*(\theta),\rho) &= (\varepsilon_\psi-3p_\psi)/m_*(\theta)\,.
\end{align}
\end{subequations}
The Fermi momentum $k_F$ and the number density $\rho$ are related as usual by \mbox{$k_F(\rho) = (3\pi^2\rho)^{1/3}$}. The scalar interactions are implied by the $\theta$-dependent fermion mass.  The $\theta$ field itself would eventually be an $r$-dependent background field, reminiscent of a mean-field approximation.
\footnote{By using the mean-field approach we are treating the scalar very much like the gravitational field. In particular, we neglect interactions associated with \eg single scalar exchanges, even if these are possible for background values of $\theta$ for which $\partial m_*(\theta)/\partial \theta \neq  0$. However, we expect this force (which may be effectively long range for light scalar masses) to be suppressed by the small effective coupling $\sim  m/M_\phi \ll 1$ and therefore irrelevant for the thermodynamic properties of the system of fermions. It is also implicitly understood that the scalar is light enough to coherently couple to the Fermi gas, \ie $m_\phi \lesssim \rho^{1/3}$.}
In later stages, changing variables to the chemical potential $\mu$ would prove helpful since it must be a continuous parameter in any static solution where chemical equilibrium is assumed. This change of variables is done by identifying the Fermi energy with the chemical potential, namely $k_F(\mu) = \sqrt{\mu^2-m_*^2(\theta)}\Theta(\mu-m_*(\theta))$. From this definition it should be understood that, for a given $\theta$, for values of $\mu$ below the mass threshold $m_*(\theta)$ the total energy and pressure of the system are $\mu$-independent and originate only from the scalar field, \ie $\varepsilon \to V(\theta)$ and  $p \to -V(\theta)$.
\\ \\
The fermion field $\psi$ describes nuclear matter in its simplest form: pure non-interacting neutrons, believed to be the main component in NSs and provide the dominant source of energy density in the non-relativistic limit. Such a description leaves out important ingredients such as additional particles (protons, electrons, muons) and interactions, namely the electroweak and nuclear forces. The latter plays a critical role, as nuclear interactions become increasingly important at high densities. All of these generate $O(1)$ corrections to the main predictions (\eg maximal mass of the bound object), which rely on the balance of all relevant forces.
Nevertheless, the pure neutron gas model is useful when extracting order-of-magnitude effects due to physics beyond the SM, and thanks to its simplicity it allows a clear identification of the physical processes behind those effects.
In addition, the formalism outlined above can be extended to different models of interacting and non-interacting Fermi gases, or adapted to the phase of matter in white dwarfs \cite{Balkin:2022qer}.\footnote{In this regard, we note that the nuclear force, mediated at leading order by pion exchange, could change in a scalarized system, via a $\theta$-dependent pion mass and interactions. We leave the detailed discussion of the impact of a sourced scalar on nuclear physics for a future publication, see for instance \cite{Ubaldi:2008nf}.}
\\ \\
The coupled system \Eq{eq::TOV} can in principle be numerically solved by specifying the initial conditions $p(0)$ and $\theta(0)$, with the remaining initial conditions $\theta'(0)=M'(0)=0$ dictated by radial symmetry. In practice, however, finding valid static solutions for \Eq{eq::gen_KG} is challenging. This can be understood by the classic intuition provided by Coleman~\cite{Coleman:1978ae}. Static solutions of the scalar EOM are analogous to classical one-dimensional trajectories in an inverted potential, where the radial direction plays the role of time. In this picture, a valid static solution is one which connects one maximum of the potential to another, with the tail of the scalar profile staying exponentially close to $\theta_0$ for arbitrarily large values of $r$. These type of trajectories are inherently chaotic, given that small variations to the initial condition would cause either an over- or an under-shoot.\footnote{One important difference w.r.t.~\cite{Coleman:1978ae} is that in our scenario the explicit radial dependence of the effective scalar potential, through the pressure dependence of $\rho_s(\theta(r),p(r))$, translates into a time-dependent potential in the classical trajectory analogy. This leads to a violation of energy conservation that could complicate the over-/under-shooting argument.  However, since pressure is always continuous, this dependence can be neglected in small regions where $p$ can be treated as constant.} Thus, viable static solutions of \Eq{eq::gen_KG}  typically require to tune the initial condition $\theta(0)$. This issue can be avoided in case \Eq{eq::gen_KG} is solved in isolation by adding a fictitious friction term~\cite{Hook:2017psm}. However, this requires neglecting the back-reaction of the scalar field on the density profile, which is precisely the effect we are after. Therefore, our numerical solutions of \Eq{eq::TOV} are based on an automatized shooting method, which tunes the value $\theta(0)$ for a fixed $p(0)$ until a viable static solution is found. 


\subsection{Equation of state: the negligible gradient limit}
\label{subsec::micro_EOS_neg_gradient}

Finding a solution to the coupled system \Eq{eq::TOV} is significantly simpler if there is a separation between the typical length scale of the scalar field $\lambda_{\phi}$ and the NS radius $R$,
\begin{align}
\lambda_{\phi} \equiv f/\Lambda^2_{\text{\tiny eff}} \ll R\,.
\label{eq::largeness}
\end{align}
We refer to this particularly simple limit as the negligible gradient limit, in which all $\theta'(r)$ and $\theta''(r)$ terms in \Eq{eq::TOV} can be neglected. The detailed derivation of this limit from a dimensional analysis of \Eq{eq::gen_KG} is presented in \App{app::dim_anal_neg_grad}. $\Lambda^2_{\text{\tiny eff}}$ is a scale typically associated with either the potential term $\sqrt{\partial V(\theta)/\partial \theta}$ or the Fermi gas term $\sqrt{\rho_s|\partial m_*(\theta)/\partial \theta|}$. For a NS this limit roughly translates into $f/\mpl{}\ll \Lambda^2_{\text{\tiny eff}}/m^2$.

In this limit, solutions to the system can be found in terms of standard thermodynamic quantities. The value of the scalar field at a given number density $\rho$ or chemical potential $\mu$ is then determined either by minimizing the total energy $\varepsilon(\theta,\rho)$ or the grand canonical potential $\Omega(\theta,\mu) \equiv \varepsilon-\mu \rho = -p(\theta,\mu)$ w.r.t. $\theta$
\begin{align}
\frac{\partial \varepsilon(\theta,\rho)}{\partial \theta}\biggr|_{\rho} 
=\frac{\partial \Omega(\theta,\mu)}{\partial \theta}\biggr|_{\mu} 
=\frac{\partial V}{\partial \theta}+ \rho_s \frac{\partial m_*(\theta)}{\partial \theta}=0\,.
\label{eq::microEOS}
\end{align}
Here $\rho_s = \partial \varepsilon_\psi/\partial m_*|_\rho = - \partial p_\psi/\partial m_*|_\mu$ depends on the chosen free variable, namely either $\rho$ or $\mu$.
\Eq{eq::microEOS} defines the microscopic EOS, along with $\partial \varepsilon/\partial \rho|_\theta = \mu$ and $\partial \Omega/\partial \mu|_\theta = \rho$.
Unsurprisingly, \Eq{eq::microEOS} is the scalar EOM in the limit where the scalar derivatives are negligible, \ie \Eq{eq::gen_KG} with its LHS set to zero. 
For \Eq{eq::microEOS} to have non-trivial solutions, i.e.~for scalarization to take place, there must be a region where the two terms appearing in it are comparable.
The condition $\rho_s {\partial m_*}/{\partial \theta}\sim{\partial V}/{\partial \theta}$ implies that the compact object must be dense enough for non-trivial solutions to exist. This condition, along with the largeness condition of \Eq{eq::largeness}, are essentially the same conditions discussed in the context of scalar sourcing at finite density in \Refs{Hook:2019pbh,Balkin:2021zfd}.

Solutions of \Eq{eq::microEOS} are of the form $\theta(\mu)$ (or $\theta(\rho)$). 
This allows us to express the total energy density and pressure of the system in terms of a single independent variable, \eg the chemical potential, such that $\varepsilon(\mu) = \varepsilon(\theta(\mu),\mu) $ and $p(\mu) = p(\theta(\mu),\mu)$.
By constructing the EOS using $\mu$ as the free parameter, the preferred phase (with maximal pressure) is always selected and the procedure outlined above produces the thermodynamically stable EOS. This ensures the continuity of $\mu$ and $p$ across a phase transition boundary, which is required for chemical and mechanical stability.\footnote{As we discuss extensively below, there can exist a meta-stable, potentially long-lived, branch of the EOS. In this case, it is necessary to use $\rho$ as the free parameter.}
At this point one can readily construct the effective EOS, \ie $\varepsilon(p)$, and numerically solve the usual TOV equations
\begin{subequations}
\label{eq::TOV_w_microEOS}
\begin{align}
& p' =- \frac{(p+\varepsilon )}{ \mpl{2}r^2 }\left( 1-\frac{2 M}{ r \mpl{2}}\right)^{-1}\left(   4 \pi   r^3 p +M\right)\,,
\\
&M'= 4 \pi r^2  \varepsilon   \,,
\end{align}
\end{subequations}
given the initial condition $M(0)=0$ and some internal pressure $p(0)$.
\\ \\
Let us get a qualitative understanding on how the effective mass of the fermion $m_*(\theta)$ would change at increased densities in light of \Eq{eq::microEOS}. As we infinitesimally increase $ \rho_s$ (\ie increasing $ \rho \simeq \rho_s$ in the non-relativistic limit), \Eq{eq::microEOS} can only be satisfied if ${\partial m_*}/{\partial \theta}<0$, namely if the mass of the fermion decreases. In other words, the increase in $V(\theta)$ due to the deviation from $\theta_0$ would be compensated by the decrease in the energy of the Fermi gas.\footnote{For concreteness, we use  $\varepsilon(\theta,\rho)$ as the relevant quantity for this particular discussion, but similar arguments can be made using $\Omega(\theta,\mu)$.} Indeed, for a fixed number density $\rho$, a Fermi gas has less energy when the mass of the fermion is decreased. Thus, we find that solutions of \Eq{eq::microEOS} always satisfy the upper bound $m_*(\theta)\leq m_*(\theta_0) \equiv m$ at all densities. It is also useful to consider the opposite regime of very high densities, where we identify two types of solutions for \Eq{eq::microEOS}. The first type is relevant if there exists a $\theta_{\infty}$ for which $m_*(\theta_{\infty})=0$. 
Then, \Eq{eq::microEOS} is solved at arbitrary high densities along a curve in the $\{\theta,\rho\}$ plane defined by
\begin{align}
\label{eq::theta_inf_case1}
\rho_s(\theta,\rho) 
= \left| \frac{{\partial V}/{\partial \theta}}{{\partial m_*(\theta)}/{\partial \theta}}\right|
_{\theta=\theta_{\infty}} 
\equiv \rho_{s,\infty}= \text{const.}
\end{align}
In the ultra-relativistic approximation $\rho_s(\theta,\rho) \sim  \rho^{2/3}m_*(\theta)$, therefore the condition above is satisfied by $m_*(\theta) \sim \rho_{s,\infty}/\rho^{2/3}$, which is achieved by taking $\theta$ close enough to $\theta_{\infty}$ for $\rho \gg \rho_{s,\infty}$.
Therefore, in this type of unbounded solution, $m_*(\theta)$ remains positive and approaches $0$ from above as the density is increased. This implies that the effective mass of the fermion can be much smaller than its zero density value $m$ at high enough densities. 

If $m_*(\theta)$ is bounded from below and does not cross $0$, we find the second type of solution: the asymptotic value of $\theta$ at high densities would then be $\theta_\infty$ for which the first derivative vanishes, namely $({\partial m_*(\theta)}/{\partial \theta})|_{\theta=\theta_\infty}=0$. This can be easily understood as the solution of \Eq{eq::microEOS} in the limit where the contribution from $\partial V(\theta)/\partial \theta$ is negligible and $\rho_s \neq 0$. 
In this type of solution, $m_*(\theta)$ remains positive and approaches $m_*(\theta_\infty)$ from above as the density is increased. Depending on the function $m_*(\theta)$, both $m_*(\theta_\infty) \lesssim m$ and $m_*(\theta_\infty) \ll m$ are possible. The scalar density $\rho_s$ increases as $\rho$ increases while the mass is fixed, which implies that $\theta(\rho)\to\theta_{\infty}$ at high densities, making this solution self-consistent. In both cases discussed above, we find that any solution of \Eq{eq::microEOS} satisfies also the lower bound $m_*(\theta)>0$ at all densities.
\\ \\
What kind of EOS can we expect? 
To answer this question, let us first discuss the qualitative effects of the scalar field. 
The first effect we can consider is the reduction of the mass of $\psi$. This has the generic effect of stiffening the EOS, which can be seen easily \eg in the non-relativistic approximation (neglecting $V(\theta)$, $p \simeq p_{\psi}$), 
\begin{align}
p_{\psi}^{\text{\tiny NR}} \propto \varepsilon_{\psi}^{\text{\tiny NR}}  \left(\frac{\varepsilon_{\psi}^{\text{\tiny NR}} }{m^4_*(\theta)}\right)^{2/3}\;\; \to \;\; \frac{\partial p_{\psi}^{\text{\tiny NR}} }{\partial m_*(\theta)}<0\,.
\end{align}
A reduction of the mass, therefore, leads to a larger pressure for a fixed energy density, \ie to a stiffer EOS that can support a larger total mass for a given radius. On the other hand, the additional contribution of $V(\theta)$ would generically lead to a softening of the EOS state. Again \eg in the non-relativistic approximation,
\begin{align}
p^{\text{\tiny NR}} \simeq 
c \, m_*^{-8/3} \big( \varepsilon^{\text{\tiny NR}}-V(\theta) \big)^{5/3}-V(\theta)\;\; \to \;\; \frac{\partial p^{\text{\tiny NR}} }{\partial V(\theta)}<0\,, \label{eq::potential_softening}
\end{align}
where $c$ is a numerical constant. An increase in $V(\theta)$ leads to smaller pressure for a fixed energy density, therefore to a softer EOS.  

These two competing effects split the parameter space of any model into two qualitatively different regions.  
First, a coexistence (CE) region, in which a phase of matter, with $\theta = \theta^{\text{\tiny CE}} \neq \theta_0$, is accessible above a certain critical pressure. 
In this case, for internal pressures above the critical pressure, the bound object can be described as a hybrid star, with a core in one phase ($\theta \simeq \theta^{\text{\tiny CE}}$) and a crust in another ($\theta \simeq \theta_0$). 
Below that critical pressure, only the low-density phase is present. 
Such phase transitions typically soften the EOS, and the resulting hybrid stars are less massive in comparison to stars made of matter in their low-density phase. 

We dub the rest of the parameter space the new ground state (NGS) region. As the name suggests, at high enough densities matter can transition to a new, stable ground state with $\theta = \theta^\ngs$. Stars could be totally stable only in this new phase, while dilute stars (with $\theta \simeq \theta_0$) can be long-lived until a fluctuation causes them to transition to the stable phase. Importantly, the EOS for the new ground state could be stiffer in comparison to the low-density (meta-stable) phase and therefore may support bound objects with larger masses. The NGS region shares some similarities with strange stars~\cite{Witten:1984rs,Farhi:1984qu,Alcock:1986hz}.
\\ \\
To better define the CE and NGS regions, let us denote our parameter space as $\vec{\alpha} = \{\alpha_i\}$, namely the space of parameters (couplings and scales) that fix the $m_*(\theta)$ and $V(\theta)$ functions, see \Eq{eq:mV}. The two regions can be easily identified in terms of preferred phases. In this case, it is useful to pick $\mu$ as the independent variable, where the preferred phase is the one with maximal pressure.

\begin{figure}
     \centering
    \includegraphics[width=0.75\textwidth]{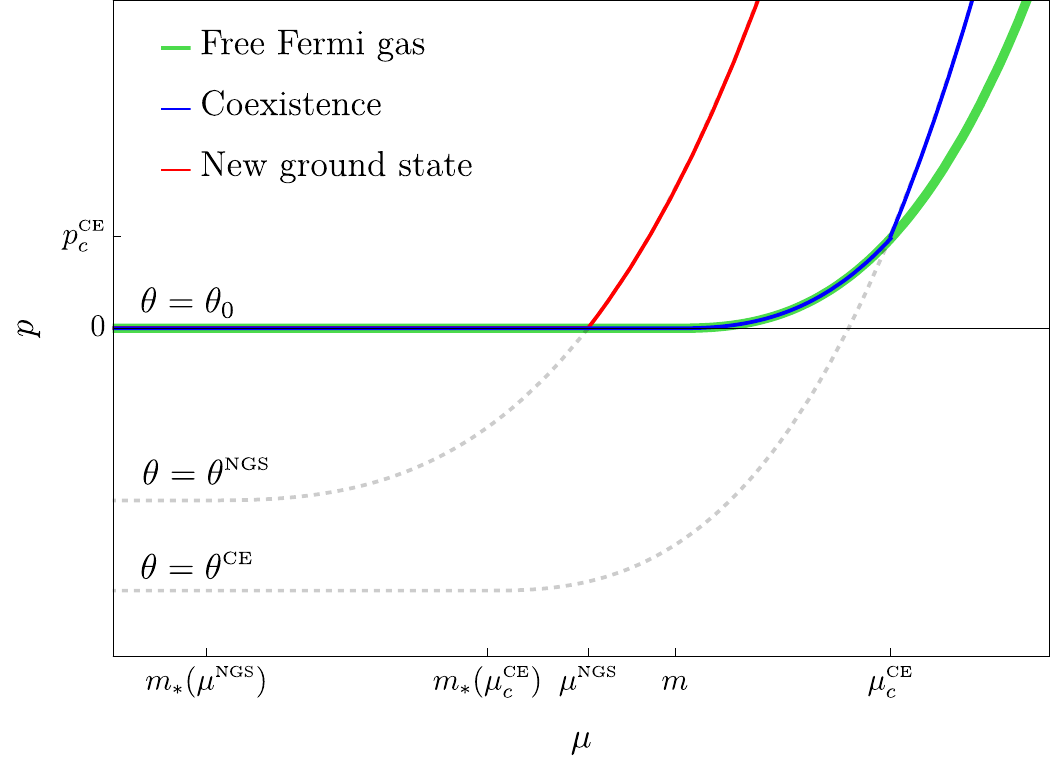}
         \caption{
         Pressure as a function of chemical potential. 
         The thick, green curve describes a free Fermi gas. 
         The blue curve describes a first-order phase transition from $\theta_0$ to some $\theta^{\text{\tiny CE}}$, typical in the CE region of parameter space. 
         The phase transition occurs at a critical chemical potential $\mu_c^{\text{\tiny CE}}$ where the pressure of both phases are equal, denoted here by $p^{\text{\tiny CE}}_c$. 
         The red curve describes the NGS, with $\theta = \theta^\ngs$ starting at the $p=0$ point at non-zero $\mu^\ngs<m$. 
         This plot demonstrates how the intersection point between the $\theta \neq \theta_0$ curves and the $\theta = \theta_0$ thick green curve, which is controlled by the properties of $m_*(\theta)$ and $V(\theta)$, determine whether a certain parameter point belongs to the CE or the NGS region.}
         \label{fig::mu_p_plot}
\end{figure}

\begin{figure}
     \centering
         \includegraphics[width=0.7\textwidth]{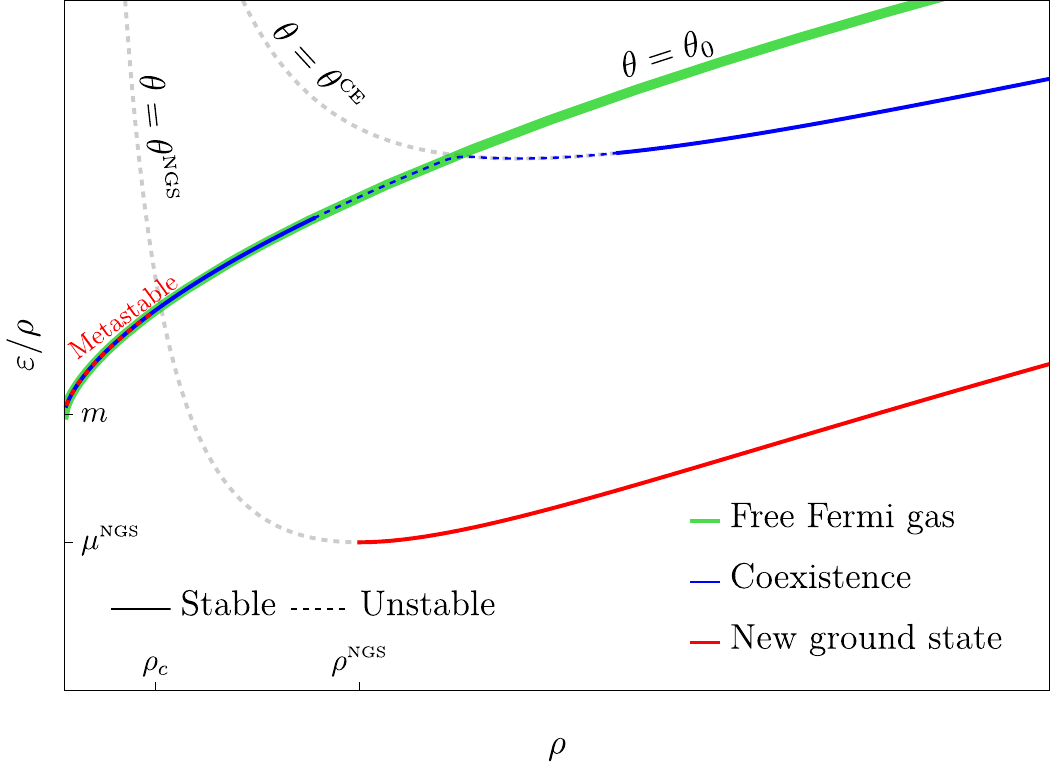}
         \caption{The binding energy $\varepsilon/\rho$ as a function of number density $\rho$. The thick, green curve describes a free Fermi gas. The blue curve describes a first-order phase transition typical in the CE region of parameter space. The transition is accompanied by a discontinuity in $\varepsilon$ and $\rho$, in the region plotted here as the dashed blue line. The phases at the edges of the dashed blue line have the same pressure. Both the thick green and the blue curves share the same ground state at $\rho \to 0$, where the binding energy is simply the rest mass $m$. The red curve describes the binding energy of the NGS. An absolutely stable branch is defined for $\rho\geq\rho^\ngs$, with the NGS at $\rho^\ngs$. A meta-stable branch equivalent to a free Fermi gas in found at $\rho<\rho_c$. The region $\rho_c<\rho<\rho^\ngs$ is completely unstable.}
         \label{fig::energy_per_particle}
\end{figure}

\begin{figure}
     \centering
         \includegraphics[width=0.7\textwidth]{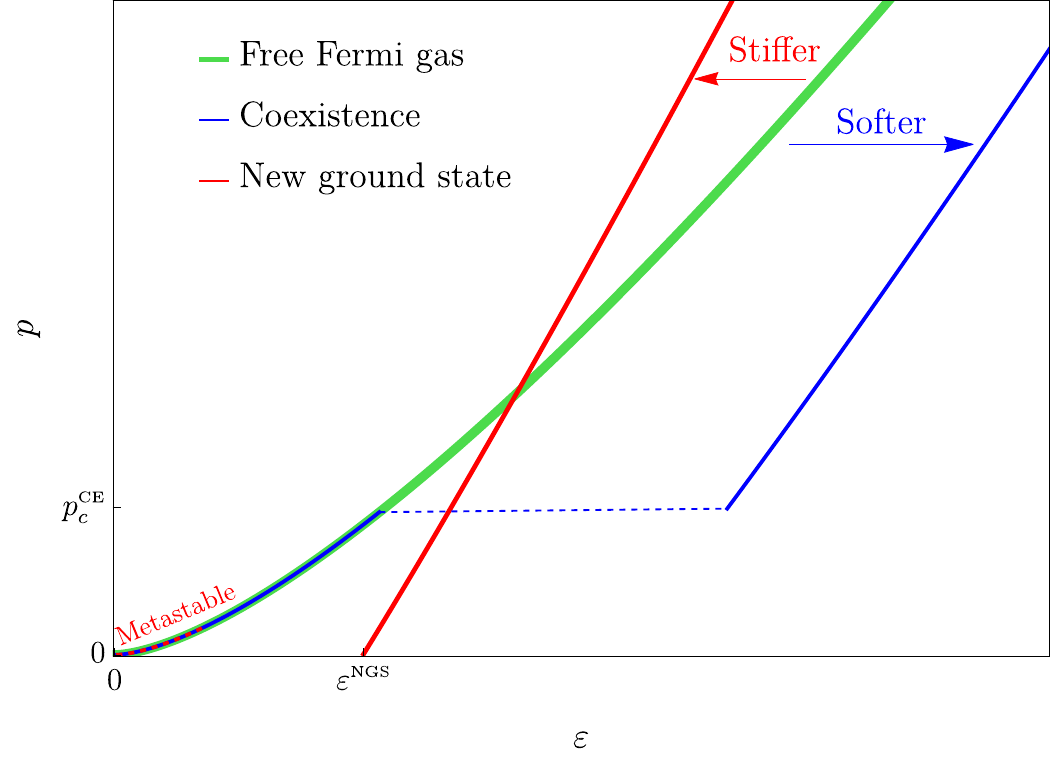}
         \caption{Pressure as a function of energy density for the various regions discussed in the text. The thick, green curve describes a free Fermi gas. The blue curve describes a first-order phase transition, typical in the CE region of parameter space. While the pressure is continuous across the phase transition, which takes place at $p_c^{\text{\tiny CE}}$, the energy density is discontinuous, shown as the dashed blue line. Such a jump leads to a softer EOS, at least in some finite region.  The red curve describes the NGS, characterized by vanishing pressure at some finite energy density $\varepsilon^{\text{\tiny NGS}}$. This EOS can be stiffer than the $\theta_0$ phase at high densities. There is typically also a meta-stable branch at low energy densities, equivalent to the free Fermi gas, shown here as the dashed red curve.}
         \label{fig::EOS_plot}
\end{figure}

\subsubsection*{Coexistence region} 

This region in parameter space is defined by $p(\theta,\mu; \vec{\alpha})<0$ for all values of $\theta \neq \theta_0$ and for $0<\mu< m$. This means that the $\theta_0$ phase is preferred around the $\mu\gtrsim m$ threshold (since by definition $p(\theta_0,\mu)=0$ in the region $0<\mu< m$). A phase transition may then occur at some critical $\mu_c^{\text{\tiny CE}}(\vec{\alpha})>m$, see the blue curve in \Fig{fig::mu_p_plot}. Although this transition is always continuous in $(p,\mu)$, it could either be smooth or first-order in $(\varepsilon,\rho)$. In the latter case, there is a discontinuity (\ie ``jump''), like the one shown in \Fig{fig::energy_per_particle}.
Phase transitions typically lead to a softening of the equation of state, see \Fig{fig::EOS_plot}. We also note that one can classify three types of possible phase transitions, depending on whether the low- and high-density phases are non-relativistic (NR) or ultra-relativistic (UR). A noticeable effect arises only when at least one of the phases is NR, such that the nucleon mass plays a role, therefore only in $\text{NR}\to\text{NR}$ and $\text{NR}\to\text{UR}$ transitions. The third possible transition, $\text{UR}\to\text{UR}$, occurs when the mass is irrelevant, and therefore changing its value does not affect the EOS. Note that the $\text{UR}\to \text{NR}$ transition is not possible; as argued above, the mass of the fermion in the high-density phase is never larger.

\subsubsection*{New ground state region}

The NGS region of parameter space is defined by demanding that there exists a solution of $p(\theta^\ngs,\mu;\vec{\alpha})=0$ at a chemical potential $\mu^\ngs(\vec{\alpha})$ satisfying $m_*(\theta^\ngs)<\mu^\ngs< m$, with $\theta^\ngs(\mu,\vec{\alpha})$ the solution of \Eq{eq::microEOS}. This is shown by the red curve in \Fig{fig::mu_p_plot}. 
In order to see that there is indeed a new ground state of the system, it is useful to switch and use the number density $\rho$ as a free parameter. The condition above implies, according to the first law of thermodynamics,
\begin{align}
p(\mu^\ngs) 
= (\rho^\ngs)^2 \frac{\partial(\varepsilon/\rho)}{\partial\rho}\biggr|_{\rho=\rho^\ngs} = \mu^\ngs \rho^\ngs - \varepsilon^\ngs=0\,,
\label{eq:pNGS}
\end{align}
where $\varepsilon^\ngs=\varepsilon(\theta^\ngs,\rho^\ngs)$ and $\rho^\ngs = k_F^3(\mu^\ngs)/3\pi^2$, with $k_F^2(\mu^\ngs) = (\mu^\ngs)^2-m_*^2(\theta^\ngs)$ and $\theta^\ngs(\vec{\alpha})$ evaluated at $\mu^\ngs$.
From this condition we learn two things,
\begin{enumerate}
\item There is a minimum of the function $\varepsilon/\rho$ at $\rho=\rho^\ngs$.
\item The value of the function at that minimum is $\frac{\varepsilon^\ngs}{\rho^\ngs} = \mu^\ngs <m$.
\end{enumerate}
This new and deeper minimum is shown in \Fig{fig::energy_per_particle}.
We find that the function describing the energy per particle, namely $\varepsilon/\rho$, has a global minimum at $\rho^\ngs$, which is lower than the minimum at $\rho=0$,
since $\lim_{\rho\to0}\varepsilon(\theta_0,\rho)/\rho = m$. This implies the existence of a new ground state for matter with $\theta^\ngs=\theta(\rho^\ngs)$. This is analogous to the effect of the nuclear force in nuclear matter. The short-distance repulsion and long-distance attraction are balanced at nuclear saturation density $\rho_0 \approx 0.15~\text{fm}^{-3}$, \ie the density of nuclei, which are the ground states of nuclear matter. In the presence of the NGS, the EOS has a stable branch that reaches $p=0$ at some non-vanishing number density $\rho^\ngs$ or energy density $\varepsilon^\ngs$, see \Fig{fig::EOS_plot}.
Importantly for our discussion, the EOS of this new phase could be stiffer than the $\theta_0$ phase, and therefore can potentially support bound objects of larger mass. Since the NGS is not continuously connected to the $\theta_0$ phase,
matter below some critical density $\rho<\rho_c$ is meta-stable, see \Fig{fig::energy_per_particle}. 
Given a system at sub-critical density, any density fluctuation large enough to overcome the potential barrier can cause a phase transition, as long as its spatial extent is large compared to $\lambda_\phi$, even if small compared to the size of the system (due to the negligible gradient limit formulated, see  \Eq{eq::largeness}). The region $\rho_c<\rho<\rho^\ngs$ corresponds to negative total pressure and is therefore completely unstable.\footnote{We should note that for $m_*(\theta)$ linear in $\theta$ ($n=1$ in \Eq{eq:mV}) there exists another (hydrodynamical) instability due to an imaginary speed of sound $c_s^2 = \partial p/\partial \varepsilon < 0$ \cite{Gao:2021fyk}. This is absent for $m_*(\theta)$ quadratic in $\theta$ ($n=2$ in \Eq{eq:mV}) since $\theta_0$ is not continuously connected to $\theta^{\text{\tiny CE}}$ (nor to $\theta^\ngs$) as one varies $\rho$.} 
\\ \\
Let us now focus on the NGS, and take the contribution coming from $V(\theta)$ to be negligible. Furthermore, let us assume that the effective fermion mass remains approximately constant in the NGS; this would be the case in models where $m_*(\theta)$ is positive and bounded from below at some $\theta=\theta_\infty$ (see discussion below \Eq{eq::theta_inf_case1}). Thus, if the total effect of the scalar interactions can be described as simply reducing the mass of the fermion to some density-independent value $m_*(\theta_\infty)<m$, the maximal mass and corresponding radius of a star composed of matter in this new phase can be easily calculated using the standard TOV equations, to find
\begin{align}
M_{\text{\tiny max}} \approx (0.7~M_{\odot} )\left(\frac{m_N}{m_*(\theta_\infty)} \right)^{2}\,, \;\;\; R(M_{\text{\tiny max}}) \approx (9.3~\text{km}) \left(\frac{m_N}{m_*(\theta_\infty)} \right)^{2}\,,
\label{eq::max_mass_estimate}
\end{align}
where we have set $m = m_N$, with $m_N \approx 939~\text{MeV}$ the neutron mass.
Clearly, a reduced fermion mass has a strong effect on the NS maximal mass, as well as on the corresponding radius. This effect is potentially much larger than the usual $O(1)$ effect one gets by using different EOSs, which model dense matter using different approaches (see \eg \cite{Lattimer:2012nd,Oertel:2016bki}).

On the other hand, the inclusion of the scalar self-interactions encoded in $V(\theta)$ would generically have the opposite effect and drive the maximal mass to lower values compared to the simple case above.
Indeed, in the regime where the reduction in the fermion mass is so large that the Fermi gas becomes ultra-relativistic, $m_*(\theta_\infty)$ ceases to play a relevant role in the EOS in comparison with (a fixed) $V(\theta_{\infty})$.
The EOS then takes a particularly simple form 
$\varepsilon  \simeq 3p+4V(\theta_{\infty})$,
and we find 
\begin{align}
\label{eq::maxMass_Vinf_scaling}
M_\text{max} \approx (1.1 M_\odot) \left(\frac{(0.2~\GeV)^4}{V(\theta_{\infty})}\right)^{1/2}\,, \;\;\;  R(M_{\text{\tiny max}}) \approx (5.8~\km) \left(\frac{(0.2~\GeV)^4}{V(\theta_{\infty})}\right)^{1/2}.
\end{align}

While the maximal mass is of particular interest given its constraining power from single NS observations, other macroscopic parameters such as compactness, $C \equiv GM/R$, are of special relevance for what regards the sensitivity of GW observatories such as LIGO, see \eg \cite{Giudice:2016zpa}. When the main effect of the scalar field is just the reduction of the neutron mass, the compactness is approximately the same as for the free Fermi gas, \ie $C \simeq 0.11$, see \Eq{eq::max_mass_estimate}. The compactness is larger if the scalar potential also plays a role, and it is maximal in the ultra-relativistic limit, where it takes the value $C \simeq 0.27$, see \Eq{eq::maxMass_Vinf_scaling}. We have explicitly checked that these two values delimit, to good approximation, the range of compactness of the heaviest NSs in the NGS, \ie
\begin{align}
\label{eq::compactness}
0.11 \lesssim 
C
\lesssim 0.27\,,
\end{align}
and that such a range is fully covered independently of the NS maximum mass, see \Fig{fig::ALP_phase_space}. Therefore, while the maximal compactness is below that required for the existence of a photon sphere, $C < 1/3$, it is larger than that of standard NSs with realistic EOSs, whose compactness does not typically exceed $C \lesssim 0.23$, see \eg \cite{Lattimer:2012nd,Lattimer:2021emm}.

An important consequence of the above discussion regards the causal bounds on the maximal NS mass and compactness. Bounds on the maximal mass of NSs, independent of their radius, were originally derived by Nauenberg and Chapline \cite{1973ApJ...179..277N} as well as by Rhoades and Ruffini \cite{Rhoades:1974fn}, based on the causality requirement that the speed of sound be smaller than the speed of light, $c_s < 1$. These bounds can be understood taking the maximally-compact EOS, namely $p = \varepsilon-\varepsilon_c$ for $\varepsilon>\varepsilon_c$, and $p=0$ otherwise, from which one finds the bound $M \lesssim 4.1 M_{\odot} \sqrt{\varepsilon_0/\varepsilon_c}$ with $\varepsilon_0 = m_N \rho_0 \approx 150 \MeV \fm^{-3}$ the nuclear saturation energy density \cite{Koranda:1996jm,Lattimer:2012nd}. Given our understanding of matter around nuclear density $\rho_0$, for $\varepsilon_c \simeq 2 \varepsilon_0$ this approximately reproduces the original bounds on the maximal NS mass, $M_{\text{\tiny max}} \simeq 3 M_{\odot}$.
This seems to be in contradiction, for example, with our simple estimate in \Eq{eq::max_mass_estimate}. However, this causal bound is clearly based on our assumed understanding of matter at $\rho \lesssim \rho_0$. In the presence of light scalars coupled to nucleons, nuclear matter as we know it in nuclei may be only metastable (see \Sec{subsec::finite_f_meta_stability}), and the NGS energy density may in fact be lower than $\varepsilon_0$, allowing for a larger maximal NS mass.

Bounds on the maximal compactness of NSs are instead not violated by stars in the NGS. These include the $\varepsilon_c$-independent causal upper bound from the maximally-compact EOS just discussed, $C \lesssim 0.35$ \cite{1984ApJ...278..364L,Haensel:1999mi,Lattimer:2012nd}, as well as Buchdahl's limit $C \leqslant 4/9$, which ensures the finiteness of the pressure inside the NS \cite{Buchdahl:1959zz,Weinberg:1972kfs}.

Another observable to consider in the context of NSs in the NGS is the rotation frequency of pulsars, which is very slowly varying in time and precisely measured, see \eg \cite{Hessels:2006ze}. 
Due to their resemblance at the macroscopic level, it is worth commenting on the rotation periods of strange stars.
In Ref.~\cite{Haensel:2009wa} it was shown that the maximal rotation frequency of strange stars can be mildly larger than standard NSs for the same masses and radii, by a factor $\lesssim 10\%$. 
The fact that strange stars have the potential to rotate faster also impacts the causal bounds on the rotation periods, which are typically slightly weaker compared to standard NSs \cite{Koranda:1996jm,Glendenning:1997wn,Haensel:1999mi}.

Besides rotation periods, it would be interesting to further explore other related observables such as the moment of inertia and the spin-induced quadrupole moment \cite{Hartle:1967he,Hartle:1968si}, see also e.g.~\cite{Glendenning:1997wn}. These observables have been studied mostly in the context of scalar-tensor modifications of gravity \cite{Pani:2014jra,Doneva:2014faa,Doneva:2016xmf,Popchev:2018fwu}. Their importance goes beyond the characterization of the macroscopic properties of NSs in that the moment of inertia and quadrupole in GR are found to be related in a way that weakly depends on the EOS \cite{Yagi:2013bca,Yagi:2013awa}. These universal relations, also known as I-Love-Q relations, also involve tidal Love numbers such as the tidal deformability \cite{Flanagan:2007ix}, which parametrizes the quadrupole response of a star to the gravitational field of a companion object.
Strange stars have been found to exhibit much larger tidal deformabilities than normal NSs \cite{Postnikov:2010yn}, yet the universal relations have been shown to hold regardless, see e.g.~\cite{Yagi:2016bkt}.
Perhaps the stronger indication that this set of observables is likely different for NSs in the NGS than for standard NSs is that departures from the I-Love-Q relations are typically significant in scalar-tensor theories, as reviewed in \cite{Doneva:2022ewd}.


\subsection{Finite gradient effects}
\label{subsec::finite_f_effects}

\subsubsection{Meta-stability}
\label{subsec::finite_f_meta_stability}

The negligible gradient limit ($f/\mpl{}\ll \Lambda^2_{\text{\tiny eff}}/m^2$) is useful and applicable when considering bound objects that are much larger than the effective wavelength of the scalar field, see \Eq{eq::largeness}. For objects whose size $R$ satisfies the opposite condition
\begin{align}
R \ll \lambda_{\phi} \,,
\end{align}
the cost in gradient energy outweighs the in-medium gain in potential energy, such that the energetically favorable configuration remains $\theta=\theta_0$.
This limit has an important implication: the new stable phase of matter is not accessible in small systems, even if these are dense enough. 
For instance, considering a low value of the scalar decay constant, \eg $f=10^3~\text{GeV}$, and taking a conservative estimate $\Lambda^2_{\text{\tiny eff}} \sim (1~\text{GeV})^2$ (which is relevant for nuclei and NSs), we find that the effective wavelength is $\lambda_{\phi}\sim 200~\text{fm}$, two orders of magnitude larger than nuclear radii $\sim1~\text{fm}\times A^{1/3}$, with mass number $A$. Since most nuclei with $A \sim O(200)$ are already short-lived, it is extremely unlikely that a nucleus with $A \sim 10^6$ is spontaneously formed in small systems. 
In practice, experimental tests typically bound $f$ to be much higher than considered above, making $\lambda_{\phi}$ much larger and the new phase even less accessible in small systems (the discussion on these model-dependent experimental tests is postponed to \Sec{sec::case_studies}).
Therefore, the stability of standard nuclei in light of the absolutely stable scalarized ground state is ensured due to the gradient energy required to displace the scalar field, and it is consistent with the fact that some nuclei are very long-lived, and in fact stable on cosmological scales. 

In addition, note that only matter fluctuations whose spatial extent is of the order of $\lambda_{\phi}$ 
can lead to a transition from the meta-stable phase to the NGS phase.
This type of large and dense regions are expected to appear first in violent events such as stellar collapse (e.g.~supernova) and stellar collisions (e.g.~binary mergers), which therefore should be the main production mechanisms of matter in its true ground state, see also e.g.~the discussion in \cite{Doneva:2022ewd}.
Depending on the critical density, the existence of a NGS can affect the formation of main sequence stars as well as stellar remnants such as white dwarfs and NSs. 
We leave the study of these effects for future investigation and limit ourselves here to the study of time-independent systems.

Finally, we note that in our simple working assumption of a free Fermi gas (of neutrons), every time a clump or nugget of matter in the NGS comes into contact with matter in the meta-stable low-density phase, it would convert it to the NGS.
In the case of strange quark matter~\cite{Witten:1984rs,Farhi:1984qu,Alcock:1986hz}, this is avoided by adding electrons according to charge neutrality. These form a cloud of size $(\alpha m_e)^{-1}$ around the strange nugget, which itself has a crust of order $\Lambda_{\text{\tiny QCD}}^{-1}$ (which is also the smallest nugget size).
The Coulomb barrier then prevents these objects from converting other nuclei to strange matter.
In our case, such a separation of scales is not necessarily present (since $\lambda_\phi \gg \Lambda_{\text{\tiny QCD}}^{-1}$) and therefore this argument is not applicable.
Nevertheless, we expect the abundance of NGS nuggets to be significantly smaller than their strange counterparts due to a suppressed production rate in extreme astrophysical environments by $\sim (\Lambda_{\text{\tiny QCD}} \lambda_\phi)^{-2}.$
Still, we believe a more careful examination of this question is warranted.

\subsubsection{Self-bound and constant density objects}
\label{sec::constant_density_objects}

For all configurations on the stable branch, the total core pressure balances two inward pressure contributions, the gravitational pressure and the pressure from the change in mass due to the scalar field, which we refer to as the gradient pressure.
We define self-bound objects (SBOs) as those for which the gravitational contribution is subdominant.
They are composed of matter in the NGS, held together by the gradient pressure of the scalar field at the boundary (or crust) of the object, which is typically of size $\lambda_{\phi}$ and where the scalar transitions from $\theta_{\infty}\rightarrow\theta_{0}$.
The opposite limit represents the conventional gravitationally-bound objects, also known as stars. As discussed below, in the NGS branch, both types of objects can have approximately constant energy density profiles.
In \App{app::object_class} we provide a detailed analytic treatment of these objects.

SBOs are typically not compact, $GM/R \ll 1$, consistent with the fact that they are well-described by the $\mpl{}\to\infty$ limit of the coupled TOV equations,
\begin{subequations}
\label{eq::TOV_selfbound_txt}
\begin{align}
& p' =-\theta' \left(\frac{\partial V}{\partial \theta}+\rho_s \frac{\partial m_*}{\partial\theta} \right) 
\,,
\label{eq::TOV_selfbound_1}
\\
& \theta''+\frac{2}{r}\theta' = \frac{1}{f^2}\left( \frac{\partial V}{\partial \theta}+\rho_s \frac{\partial m_*(\theta)}{\partial \theta}\right)\,.
\label{eq::TOV_selfbound_2}
\end{align}
\end{subequations}
Note that the pressure profile is non-trivial only in regions where $\theta' \neq 0$ and the condition in \Eq{eq::microEOS}, which defined our microscopic EOS, is not satisfied.
The smallest possible SBOs, also known as nuggets, are of size
\begin{align}
R_{\text{\tiny SBO}}^{\text{\tiny min}} \sim \lambda_\phi \,.
\end{align} 
These are the densest of all the SBOs, since for higher densities \Eq{eq::TOV_selfbound_txt} no longer admits stable solutions.
In the $R \gg \lambda_{\phi}$ limit, the interior of SBOs is well-described by constant internal pressure and energy density, as well as a constant scalar field value.
They are held together by the gradient pressure exerted in a small region of size $\lambda_\phi \ll R$, at the edge of the object.
For even lower 
densities, the SBOs become large, reaching the point where gravity can no longer be neglected.
Depending on the region of parameter space, SBOs can range many order of magnitude in size $R_{\text{\tiny SBO}}^{\text{\tiny max}} \gg R_{\text{\tiny SBO}}^{\text{\tiny min}}$. In particular, they can be as large as NSs and with similar masses. On the other hand, for large values of $f$, SBOs may not be viable configurations at all. Indeed, the existence of SBOs requires that $  R_{\text{\tiny SBO}}^{\text{\tiny max}}  >  R_{\text{\tiny SBO}}^{\text{\tiny min}} \sim \lambda_\phi $, which reduces to an upper bound on $f$ 
\begin{align}
f \lesssim \frac{\dm }{1-\dm}\mpl{} \,,
\end{align}
where $\dm=1-m_*(\theta_\infty)/m$.
For objects of size $R \gtrsim R^{\text{\tiny SBO}}_{\text{\tiny max}}$, the gradient pressure and the gravitational pressure are equally important and the full coupled set of equations of \Eq{eq::TOV} must be solved, as discussed just above \Sec{subsec::micro_EOS_neg_gradient}. 
\\ \\
As the objects in the NGS become more compact (larger and more massive, since $\varepsilon$ is approximately constant) gravity can no longer be neglected and eventually dominates. 
We identify then two distinct limits, depending on the model parameters $\vec{\alpha}$, here encoded in $\dm$ and $m_*$ (evaluated at $\theta_\infty$), $f$, and $\varepsilon^\ngs (\vec{\alpha})$.  
In the limit $\dm^{1/2}(m_*^4/\varepsilon^\ngs)(f/\mpl{}) \ll 1$, in which SBOs have energy densities close to that of the ground state, $\varepsilon^\ngs$ (where the EOS is $p \simeq 0$), the smallest gravitationally-bound objects can still be approximated as constant energy density systems. 
However, their pressure drops away from the core (as opposed to the constant pressure SBOs), at sufficiently low pressures the EOS is probed only in regions where $\varepsilon \simeq \varepsilon^\ngs \simeq \text{const}$.
At large enough core pressures, the approximation of constant energy density breaks down, and any further increase in core pressure leads to a decrease in radius, which is the typical behavior of stars described by a Fermi gas.

On the other hand, for $\dm^{1/2}(m_*^4/\varepsilon^\ngs)(f/\mpl{}) \gg 1$, the most massive and largest SBOs have $\varepsilon \gg \varepsilon^\ngs$ and already probe, by definition, the part of the EOS in which any further increase in pressure leads to stars that are no longer constant density objects. Therefore, when the gravitational pressure becomes relevant, these objects have both decreasing pressure and energy density away from their core.

Following \App{app::object_class}, we find that the maximal radius for constant energy density objects in the $\dm \ll 1$ limit is approximately given by
\begin{align}
R^{\text{\tiny max}}_{\text{const}}
\sim \begin{cases}
\frac{\mpl{}}{m_*^{8/6}(\varepsilon^\ngs)^{1/6}}  \;\;\;\;\;\;\;\;\;\; & \dm^{1/2} \left( \frac{m_*^4}{\varepsilon^\ngs} \right) \left(\frac{f}{\mpl{}}\right) \ll 1
\\
\frac{\mpl{7/6}}{\dm^{1/12} m_*^2 f^{1/6}}  &  \dm^{1/2} \left( \frac{m_*^4}{\varepsilon^\ngs} \right) \left(\frac{f}{\mpl{}}\right) \gg 1
\end{cases}\,.
\label{eq::size_of_constant_objects}
\end{align}
\\ \\
To conclude, let us note that a simple prediction regarding the mass-radius relation can be given for objects in the NGS with densities of order $\rho^\ngs$; this is the region in the EOS where the energy density becomes almost constant as $p\to0$, approaching the critical value $\varepsilon \to \varepsilon^\ngs$.
The mass of these constant-density objects is then given simply by the product of the energy density times the volume 
\begin{align}
\label{eq::MR_relation_constant_density}
M/R^3 \simeq \frac{4\pi}{3} \varepsilon^\ngs \,. 
\end{align}
Their compactness, scaling as $C \propto R^2$, is much smaller than for stars in the NGS, since their radius is much smaller. As we have explained above, as the radius grows the gravitational pressure becomes important and the energy density acquires a non-trivial profile, and one returns to the predictions given in \Eq{eq::compactness}.


\section{Case studies}
\label{sec::case_studies}


\subsection{Bounded $m_*(\theta)$ solutions}
\label{sec::bounded_models}

We begin the analysis with models in which $m_*(\theta)$ is bounded from below and does not cross zero, such that the high-density value of the scalar is $\theta_{\infty}$ defined by $(\partial m_*/\partial \theta) |_{\theta=\theta_{\infty}}=0$. We analyze these models in the negligible gradient limit, with the additional assumption that $\theta$ jumps from the trivial phase $\theta=\theta_0$ to the $\theta=\theta_{\infty}$ phase. This is a good approximation if $(\partial m_*/\partial \theta) |_{\theta=\theta_0}$, as well as $(\partial V/\partial \theta) |_{\theta=\theta_{\infty}}$, are sufficiently small. 
The first of these conditions has to be fulfilled if the model is not to be excluded by fifth force experiments, and effectively corresponds to taking $n=2$ in \Eq{eq:mV}.
The second condition is only necessary for our model-independent treatment and can easily be violated in explicit models.
Then, under these assumptions, the parameter space is reduced to simply 
$\{m_*(\theta_{\infty}),V(\theta_{\infty})\}$. As described in \Sec{sec:model}, the two phases could either coexist, leading to hybrid star configurations or describe a meta-stable phase and a new absolutely stable phase of matter. The boundary line between the CE region and the NGS region in the $\{m_*(\theta_{\infty}),V(\theta_{\infty})\}$ plane is given by
\begin{align}
p(\theta_{\infty},\mu=m)= p(\theta_0,\mu=m)=0\,.
\label{eq::ALP_NGS_region}
\end{align}
In the CE region, the two phases meet at some critical chemical potential $\mu_c^{\text{\tiny CE}}$ at equal pressures. 
In contrast, in the NGS region, it is more convenient to find the meta-stable and stable branches using $\rho$ as a free parameter. 
The critical number density $\rho_{\text{c}}$, which determines the edge of the homogeneous (meta-stable) branch in the CE (NGS) region,%
\footnote{In general, the homogeneous branch in the CE region also includes a meta-stable part which probes the $\theta=\theta_0$ EOS above the critical pressure determined by the Maxwell construction, $p^{\text{\tiny CE}}_c$, but below the critical number density.}
is defined by the scalar density in which the second derivative of the effective scalar potential flips sign,
\begin{align}
\rho_s(\theta_0,\rho_{\text{c}}) = - \left(\frac{\partial^2 V/\partial \theta^2}{\partial^2 m_*/\partial \theta^2}\right)\biggr|_{\theta=\theta_0} \,,
\label{eq:rhoc}
\end{align}
where we have expressed the scalar density in terms of the number density.

\begin{figure}
\centering        
\includegraphics[width=0.99\textwidth]{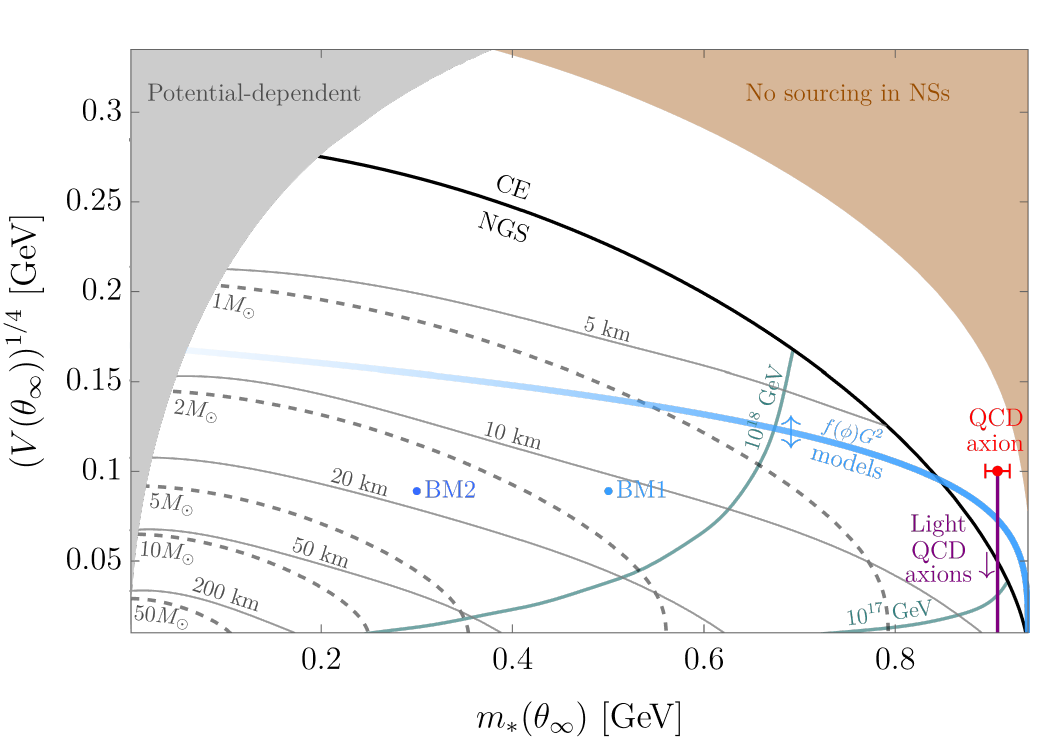}
\caption{The $\{m_*(\theta_{\infty}),V(\theta_{\infty})\}$ parameter space. The black line is the CE-NGS boundary, \Eq{eq::ALP_NGS_region}. Curves of constant stellar mass and radius for the most massive configuration allowed by the EOS are shown in dashed and solid gray, respectively. In the gray region, the $\theta=\theta_{\infty}$ approximation breaks down. In the brown region, neutron stars are not dense enough to source the scalar.
We also show where different models that conform to our simplified treatment lie in this plane: the QCD axion (red dot), see \Sec{sec::QCDaxion}; light QCD axions (purple line), \Sec{sec::lightQCDaxion}; and two generic axion benchmark points (blue dots), \Sec{sec::ALP_model}, with $m_*(\theta_{\infty})=m_N/2$ and $V({\theta_{\infty}})=2\times\left(0.075~\text{GeV}\right)^4$ (BM1) and $m_*(\theta_{\infty})=m_N/3$ and $V({\theta_{\infty}})=2\times\left(0.075~\text{GeV}\right)^4$ (BM2).
The thick-blue line describes a UV completion from a $f(\phi)GG$ interaction that allows for large nucleon couplings. As indicated by the arrows, this model can populate most of the parameter space although with decreasing calculability towards small $m_*(\theta_{\infty})$, which we indicate with a decreasing opacity. See \App{app::fGG} for more details.
Below the green contours, finite gradient effects become important for the corresponding value of $f$ shown.}
\label{fig::ALP_phase_space}
\end{figure} 

The total mass $M$ and radius $R$ of the stars are found by solving \Eq{eq::TOV_w_microEOS} numerically and scanning over the parameter space.
 In 
 \Fig{fig::ALP_phase_space}, we show the contours of constant mass and radius for the most massive stars in the NGS region. 
For small values of $V(\theta_{\infty})$, the constant $M$ and $R$ contours are approximately $V(\theta_{\infty})$-independent. The increase in mass and radius of the NGS stars are solely due to a decreased fermion mass. We recover the same scaling as in \Eq{eq::max_mass_estimate}.
On the other hand, near the gray region in (the upper-left part of) \Fig{fig::ALP_phase_space}, the NGS becomes ultra-relativistic, thus the EOS is nearly $m_*(\theta_\infty)$-independent, and we recover the scaling of \Eq{eq::maxMass_Vinf_scaling}. 
As expected, larger values of $V(\theta_{\infty})$ make the EOS softer, resulting in lighter stars. 

Let us comment on the gray region in \Fig{fig::ALP_phase_space}. It indicates where we expect the $\theta=\theta_{\infty}$ approximation to break down. This happens when the scalar density in the $\theta=\theta_{\infty}$ phase, $\rho_s(\theta_{\infty},\mu)$, drops below a value given by 
\begin{align}
    \rho_s^{\infty} = - \left(\frac{\partial V/\partial \theta}{\partial m_*/\partial \theta}\right)\biggr|_{\theta=\theta_{\infty}} \,.
    \label{eq::crit_rhos}
\end{align}
The exact position of the boundary of the gray region is in general model-dependent.
That plotted in \Fig{fig::ALP_phase_space} corresponds to models with $m-m_*(\theta) \propto V(\theta)$ (\eg the QCD axion and its lighter variations, see \Secs{sec::QCDaxion}{sec::lightQCDaxion}), in which $\rho_s^{\infty} = V(\theta_{\infty})/(m-m_*(\theta_\infty))$.
In general, this region in which $\theta$ does not jump all the way to the point where $\partial m_*/\partial \theta=0$ depends strongly on the shape of $V(\theta)$ and $m_*(\theta)$ and it has to be determined numerically.

Let us also discuss the validity of the negligible gradient limit, which we have assumed to hold in the discussion above.
Finite gradient effects become important as soon as $\lambda_\phi \sim R$, where recall $\lambda_\phi \equiv f/\Lambda^2_{\text{\tiny eff}}$, see \Eq{eq::largeness}.
At high densities, we have $\Lambda^4_{\text{\tiny eff}} = - \rho_s (\partial m_*/\partial \theta)$ (see \Eq{eq::wavelengthin} and the associated discussion), which we can approximate as $\rho^\ngs [m-m_*(\theta_\infty)]$.
As a result, most of the parameter space in \Fig{fig::ALP_phase_space} is valid for $f \ll (10^{17}-10^{18})~$GeV, except for the lower-right corner, as indicated by the green lines. As we show below, finite gradient effects typically suppress the deviations from the ideal Fermi gas, thus the results of \Fig{fig::ALP_phase_space} represent the maximal effect one can expect from a sourced scalar.

\begin{figure}
     \centering 
     \includegraphics[height=4.8cm]{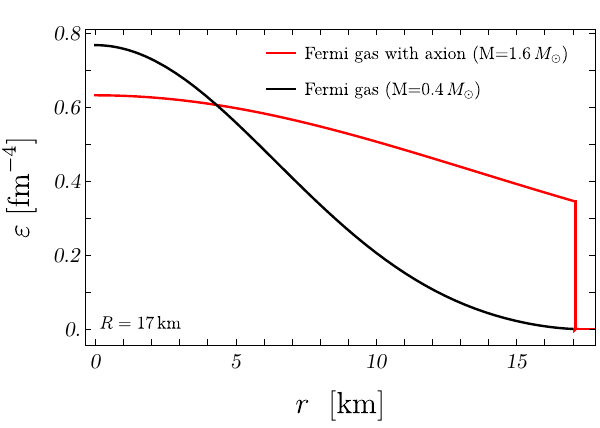} \includegraphics[height=4.8cm]{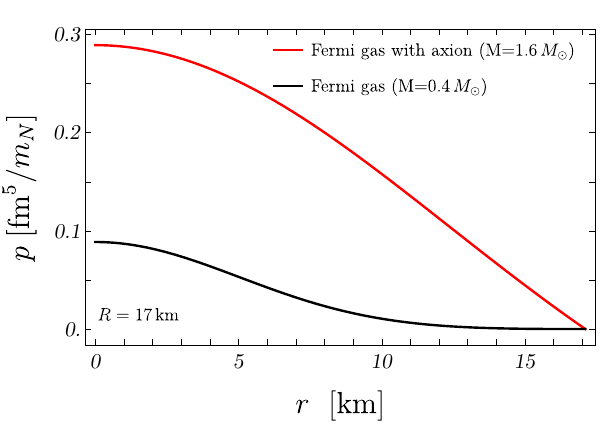}
     \\
     \includegraphics[height=4.8cm]{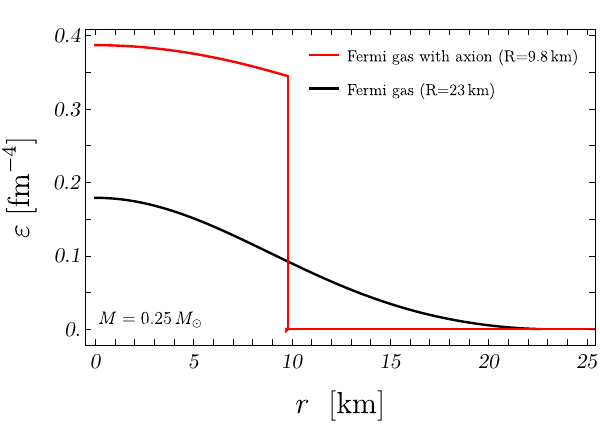} \includegraphics[height=4.8cm]{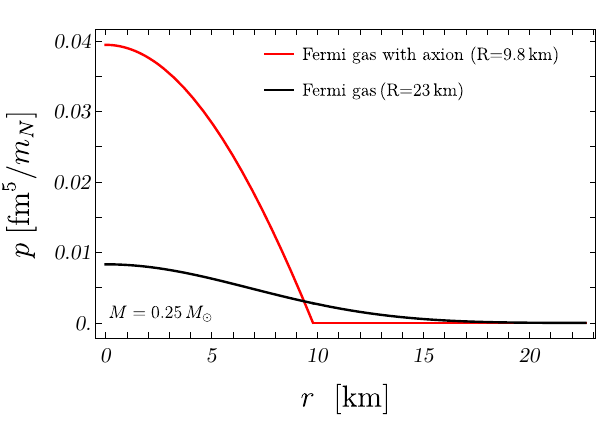}
     \caption{Top panel: Energy density (left) and pressure (right) profiles of solutions to the TOV equations. In red a configuration in the BM2 axion model with $R = 17 \km$ and $M=1.6\,M_{\odot}$, 
     and in black the free Fermi gas solution of equal radius and $M=0.4\,M_{\odot}$. Bottom panel: Profiles for stars of equal mass, $M=0.25M_{\odot}$, and $R = 9.8 \km$ and $R = 23 \km$ for the BM2 benchmark and the free Fermi gas, respectively.}
     \label{fig::profiles}
\end{figure}

Before moving to the discussion of specifics models with bounded $m_*(\theta)$, it is worth to illustrate already here the effects of a sourced scalar field on the configuration of stars in the NGS. 
In \Fig{fig::profiles} we show in red the energy density and pressure profiles of a representative point in the BM2 axion benchmark (see \Fig{fig::ALP_phase_space}), and compare it with the free Fermi gas solution, in black, of either similar radius (top panel) or similar mass (bottom panel). The emergence of the NGS can be deduced from the behavior of the energy density at the edge of the star, where it does not vanish even though the pressure does. 
We do not show the scalar profile since for this benchmark $\lambda_\phi \ll R$ and therefore the transition from $\theta_\infty$ to $\theta_0$ is very narrowly localized at the edge of the star.

\subsubsection{The QCD axion}
\label{sec::QCDaxion}

The QCD axion is an elegant solution to the strong CP problem where the QCD $\bar{\theta}$-angle is promoted to a dynamical field. The relevant part of the QCD-axion Lagrangian at energies above the QCD scale is
\begin{equation}
    \mathcal{L}_{\phi}\supset\frac{1}{2}(\partial \phi)^2+\frac{g_s^2}{32\pi^2}\frac{\phi}{f_a}G_{\mu\nu}\tilde{G}^{\mu\nu}.
    \label{eq:axionL}
\end{equation}
Here $g_s$ is the strong coupling constant and $G_{\mu\nu}$ the gluon field strength. At energies below the QCD confinement scale, a potential for the QCD axion is generated by non-perturbative effects, 
\begin{equation}
\label{eq:QCDV}
V(\phi)=-m_{\pi}^2f_{\pi}^2\left(\sqrt{1-\zud \sin^{2}\left(\frac{\phi}{2f_a}\right)}-1\right) \,,
\end{equation}
where $\zud = 4m_um_d/(m_u+m_d)^2$.
In vacuum, this potential is minimized at $\left\langle \phi  \right\rangle = 0$, such that CP is conserved, thus solving the strong CP problem.

A non-derivative coupling of the QCD axion to nucleons (neutrons and protons) is also generated at low energies, giving rise to a scalar-dependent fermion mass, see \eg \cite{Ubaldi:2008nf,Balkin:2020dsr}
\begin{equation}
\label{eq:QCDm}
m_*(\phi) = m_N + \sigma_{\pi N} \left(\sqrt{1-\zud \sin^{2}\left(\frac{\phi}{2f_a}\right)}-1\right)\,,
\end{equation}
where $\sigma_{\pi N} \approx 50 \MeV$ is the so-called nucleon sigma term, and we neglected isospin-violating contributions.

Following our definitions in \Eq{eq:mV}, the QCD axion belongs to the case with $n = 2$, \ie a quadratically coupled scalar, and we can identify 
\begin{equation}
\mphi{2} = \frac{2 m_N}{\zud \sigma_{\pi N}} f_a^2 \,, \;\;\; 
m_\phi^2 = \frac{\zud}{4} \frac{m_{\pi}^2f_{\pi}^2}{f_a^2}\,, \;\;\; 
F_\phi = f_\phi = f_a \,.
\label{eq:QCDparam}
\end{equation}
Given that $\mphi{} \gg F_\phi, f_\phi$, higher-order terms in $\phi$ are generically relevant. In addition, it is natural to identify the characteristic scale of the scalar with the axion decay constant $f_a$, therefore $f \equiv f_a$ and $\theta \equiv \phi/f_a$. The in-vacuum value of the scalar is then $\theta_0 = 0$.

\begin{figure}
	\centering 
	\includegraphics[height=7.5cm]{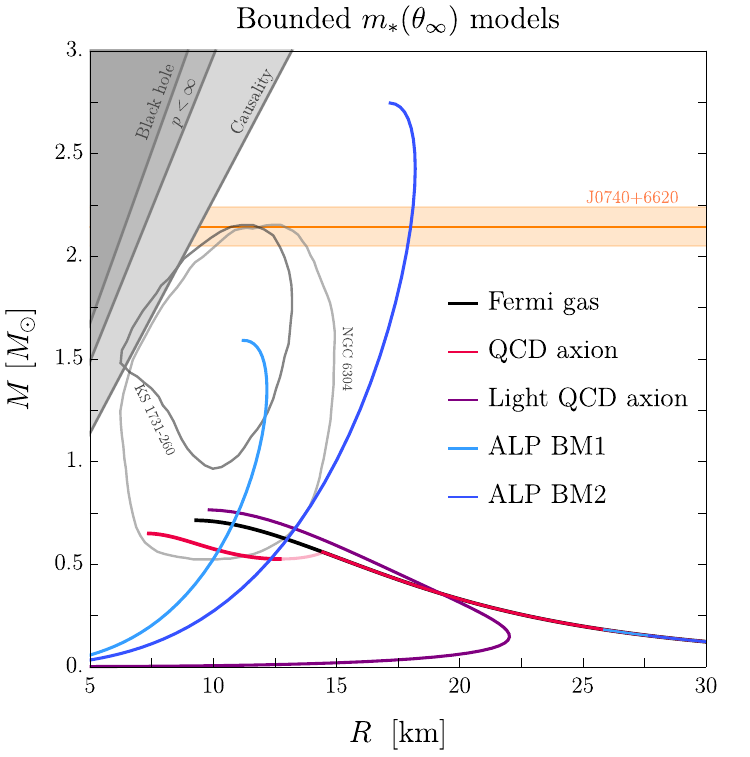} \includegraphics[height=7.5cm]{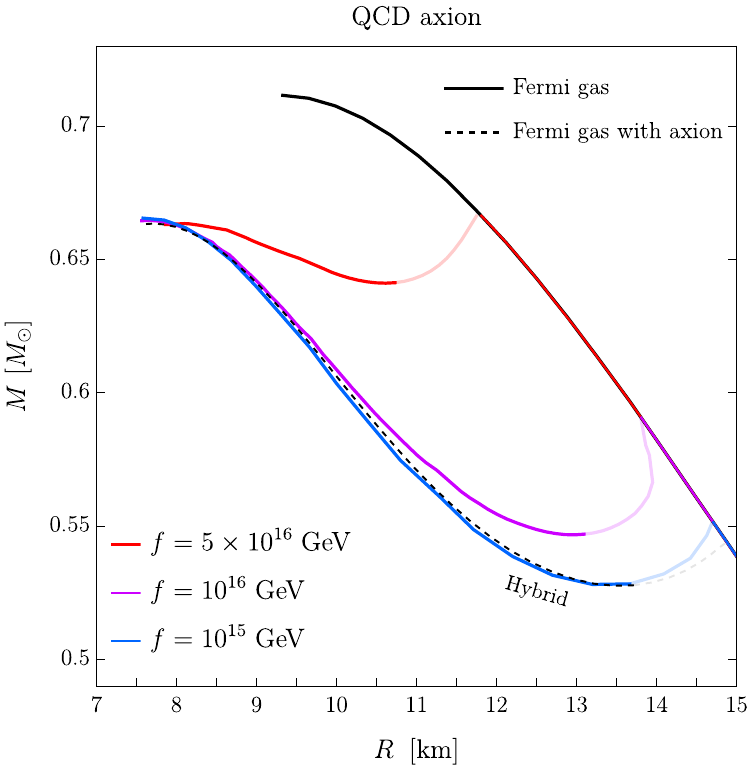}
	\caption{Left panel: $M$-$R$ curves in the negligible gradient limit for several benchmark cases as indicated in the legend, with $\epsilon=10^{-4}$ for the light QCD axion. Also shown two representative constraints at the 68\% confidence level on NS masses and radii, obtained from low-mass X-ray binaries during quiescence (NGC 6304) and thermonuclear bursts (KS 1731-260), taken from~\cite{Ozel:2016oaf}. In orange we plot the mass measurement of the millisecond pulsar J0740+6620, taken from~\cite{NANOGrav:2019jur}. The gray regions are the theoretically excluded regions with $C = GM/R \leqslant 1/2$ (Black hole), Buchdahl's limit $C \leqslant 4/9$ ($p<\infty$), and $C \lesssim 0.35$ (Causality; see the discussion below \Eq{eq::compactness}).
	Right panel: $M$-$R$ curves for the QCD axion. The free Fermi gas without the axion (solid black), the negligible gradient limit (dashed black), and including finite gradient effects for $f=\{5\times10^{16},10^{16},10^{15}\} \GeV$ in (solid) red, purple, and blue respectively. The light-colored curves are unstable configurations.}
	\label{fig::ALP_finite_f_MR_QCD}
\end{figure}

At finite baryon density, one can identify an effective in-medium QCD axion potential, which at leading order in chiral perturbation theory reads \cite{Hook:2017psm,Balkin:2020dsr}
\begin{equation}
\label{eq:QCDlowdensity}
V_{\text{\tiny eff}}(\phi,\rho_s) = V(\phi) + \rho_s \left[ m_*(\phi) - m_N \right] = \left(1 - \frac{\sigma_{\pi N} \rho_s}{m_{\pi}^2f_{\pi}^2} \right) V(\phi) \,.
\end{equation}
This potential is only valid at low densities, below around nuclear saturation, $\rho_s\simeq\rho \lesssim \rho_0$, while the exact form of the potential at high densities is unknown. This means that, for the QCD axion, keeping only the low-density potential is not realistic. Indeed, the critical density for scalarization \Eq{eq:rhoc}, where $\theta = \theta_0$ becomes unstable, would be given by
\begin{equation}
\rho_s(0,\rho_{\text{c}}) = \frac{m_\pi^2 f_\pi^2}{\sigma_{\pi N}} \,,
\label{eq:rhocQCD}
\end{equation}
which implies $\rho_{\text{c}} \gg \rho_0$ and therefore beyond perturbative control. 
In addition, at leading chiral order the pions become massless at such densities, clearly invalidating our treatment of the axion.
Nevertheless, it has been hypothesized in \cite{Balkin:2020dsr} that the sourcing of the QCD axion could be triggered by Kaon condensation. The latter is a possibility widely considered in the literature, particularly concerning the so-called hyperon puzzle, see \eg \cite{Bombaci:2016xzl}.
For the rest of this section we assume that our simple treatment based on an axion-dependent nucleon mass given by \Eq{eq:QCDm}, and the corresponding effective potential in \Eq{eq:QCDlowdensity}, hold at all relevant densities, keeping in mind that the high-density dynamics of QCD has a critical impact on the possibility that the axion actually leads to scalarized NSs.
What we learn is still useful, since for a lighter version of the QCD axion, discussed in \Sec{sec::lightQCDaxion}, the relevant densities are much lower and therefore under perturbative control.
\\ \\
In the negligible gradient limit, the system is minimized at $\theta_{\infty}=\pi$ for densities $\rho > \rho_{\text{c}}$, which implies
\begin{equation}
\label{eq:QCDAxionParam}
m_*(\theta_\infty) = m_N-\sigma_{\pi N}\left(1-\sqrt{1-\zud} \right)\,, \quad
V(\theta_\infty) = m_{\pi}^2f_{\pi}^2\left(1- \sqrt{1-\zud}\right)\,.
\end{equation}
Recall that in this case, the high-density value of the scalar corresponds to $\partial m_*/\partial \theta = 0$, which sets $\vev{\phi} = \pi f \ll \mphi{}$. This constitutes therefore an example of screened scalarization, where the screening is due to higher-order terms in the interaction of the scalar with matter.
NSs with a sourced QCD axion belong to the CE region only, see \Fig{fig::ALP_phase_space}. This is because the chemical potential corresponding to the critical density is much larger than $m_N$, see the discussion in \Sec{subsec::micro_EOS_neg_gradient}.
Since the QCD axion with $\lambda_\phi \ll R$ is compatible with all the assumptions discussed at the beginning of \Sec{sec::bounded_models}, its $M$-$R$ curve can be calculated given the values of $\{m_*(\theta_\infty),V(\theta_\infty)\}$, see the left panel of \Fig{fig::ALP_finite_f_MR_QCD} (red curve).

Moving beyond the negligible gradient limit, we show the resulting $M$-$R$ curves, found by solving the full coupled system of \Eq{eq::TOV}, in the right panel of \Fig{fig::ALP_finite_f_MR_QCD}, for different values of $f$. As expected, the phase transition from $\theta_0 = 0$ to $\theta_{\infty} = \pi$ leads to a softening of the EOS, and therefore to less massive stars. The larger stars follow the free Fermi gas line, as their densities are sub-critical, \ie $\rho(r=0) < \rho_{\text{c}}$. Smaller and denser configurations are hybrid stars, composed of a core in the $\theta_{\infty}$ phase and an exterior region in the $\theta_0$ phase.
The effect of the gradient can be recognized by two particular features.
First, the inner core of the homogeneous stars (on the right branch) can have a region that is above the critical density, as long as this region is smaller than the effective in-medium wavelength of the scalar field $\lambda_\phi \propto f$. Higher values of $f$ trace the Fermi gas line (solid black) until smaller radii and larger masses, at which point a large enough central region is created and it is energetically favorable for the high-density phase to form.
Second, for high values of $f$ the start of the hybrid branch consists of configurations where the axion is not fully sourced, \ie the value of the field does not reach $\theta_{\infty}$ at the core, thus forming a thick-wall bubble. In such configurations, the transition region of the scalar inside the hybrid star occupies a significant fraction of the whole object,  which modifies its equilibrium configuration. 
This explains the visible deviations of the finite $f$ hybrid branches (solid colored lines) compared to the negligible gradient limit (dashed line). However at higher internal pressures, once the QCD axion is fully sourced, indicating a thin-wall bubble, all the curves in the $M$-$R$ plane converge, particularly to a similar maximal mass configuration. This is not a surprise, since the existence of a thin wall is tantamount to a negligible gradient, $\lambda_\phi \ll R$.

\subsubsection{Lighter QCD axions}
\label{sec::lightQCDaxion}

The QCD axion solution to the strong CP problem has been recently extended by relaxing the relation between the potential of the axion and its coupling to the SM \cite{Hook:2017psm,Hook:2018jle,DiLuzio:2021pxd}. These QCD axions are lighter than usual, a fact that gives rise to novel astrophysical signatures~\cite{Hook:2017psm,DiLuzio:2021gos,Balkin:2022qer,Zhang:2021mks}.

For our purposes, the only yet key difference w.r.t.~the standard QCD axion is a suppressed (in-vacuum) axion potential, which can be parametrized as
\begin{equation}
    V(\phi)=- \epsilon m_{\pi}^2f_{\pi}^2\left(\sqrt{1-\zud \sin^{2}\left(\frac{\phi}{2 f_a}\right)}-1\right)\,,
\end{equation}
with $\epsilon < 1$. 
Conversely, the axion-dependent nucleon mass remains the same, see \Eq{eq:QCDm}.
Therefore, this model can be characterized by the same scales as the QCD axion, \Eq{eq:QCDparam}, except for the scalar mass, which now reads
\begin{equation}
m_\phi^2 = \epsilon \frac{\zud}{4} \frac{m_{\pi}^2f_{\pi}^2}{f_a^2}\,.
\end{equation}
As a consequence, the effective potential at finite density is given by
\begin{equation}
\label{eq:light_QCDlowdensity}
V_{\text{\tiny eff}}(\phi,\rho_s) = \left(1 - \frac{\sigma_{\pi N} \rho_s}{\epsilon m_{\pi}^2f_{\pi}^2} \right) V(\phi) \,.
\end{equation}
Due to the relative enhancement of the finite density corrections, the critical density where $\theta_0$ is no longer a minimum is $\epsilon$-suppressed compared to that of the QCD axion,
\begin{equation}
\rho_s(0,\rho_{\text{c}}) = \epsilon \frac{m_\pi^2 f_\pi^2}{\sigma_{\pi N}} \,.
\label{eq:light_rhocQCD}
\end{equation}
This is crucial since, as opposed to the QCD axion, for small enough $\epsilon$ the transition to the $\theta_{\infty}$ phase may occur at such low densities that chiral perturbation theory is valid and the pions are heavy ($\rho_{\text{c}} < \rho_0$), making our treatment in terms of a scalar-dependent nucleon mass and \Eq{eq:light_QCDlowdensity} a viable approximation.
In that case, we can reliably infer that in the ``high-density'' regime, $\rho > \rho_{\text{c}}$, $\partial m_*/\partial \theta = 0$ sets $\theta_{\infty} = \pi$ in the negligible gradient limit. While this leads to the same $m_*(\theta_\infty)$ as for the QCD axion, $V(\theta_\infty)$ is $\epsilon$-suppressed,
\begin{equation}
\label{eq:light_QCDAxionParam}
m_*(\theta_\infty) = m_N-\sigma_{\pi N}\left(1-\sqrt{1-\zud} \right)\,, \quad
V(\theta_\infty) = \epsilon m_{\pi}^2f_{\pi}^2\left(1- \sqrt{1-\zud}\right) \,.
\end{equation}
Because of the smaller potential in the scalarized phase, the NGS is now accessible if $\epsilon$ is small enough.
Since all the parameters in \Eq{eq:light_QCDAxionParam} except for $\epsilon$ are fixed experimentally ($\zud \approx 0.88$, $m_\pi \approx 135 \MeV$, $f_\pi \approx 92 \MeV$), one can phrase the condition for the existence of a NGS, see \Eq{eq::ALP_NGS_region}, as the following upper bound
\begin{equation}
\label{eq:epsilonNGS}
\epsilon \lesssim \frac{4 \sqrt{2}}{15 \pi^2} \frac{m_N^4}{m_\pi^2 f_\pi^2} \left( \frac{\sigma_{\pi N}}{m_N} \right)^{5/2} 
\left(1-\sqrt{1-\zud} \right)^{3/2} \approx 0.07
\,.
\end{equation}
at leading order in $\sigma_{\pi N}/m_N$ (in the non-relativistic limit).
The number density of the axionic (absolutely stable) ground state is given, under the same approximations, by
\begin{equation}
\label{eq:rhoepsilonNGS}
\rho^\ngs \sim \left[ m_*(\theta_\infty) V(\theta_\infty) \right]^{3/5}
\sim \left( \epsilon m_N m_\pi^2 f_\pi^2 \right)^{3/5} \,.
\end{equation}
Since the light QCD axion with $\lambda_\phi \ll R$ is compatible with all the assumptions discussed at the beginning of \Sec{sec::bounded_models}, its $M$-$R$ curve can be calculated given the values of $\{m_*(\theta_\infty),V(\theta_\infty)\}$, see the left panel of \Fig{fig::ALP_finite_f_MR_QCD} (purple curve).
As expected, since the EOS of the NGS is slightly stiffer, it leads to slightly more massive stars than the free Fermi gas, enhancing the maximal mass of NSs by $(m_N/m_*(\theta_\infty))^2-1 \approx 10\%$.

\begin{figure}[t]
	\centering    
	\includegraphics[height=7.5cm]{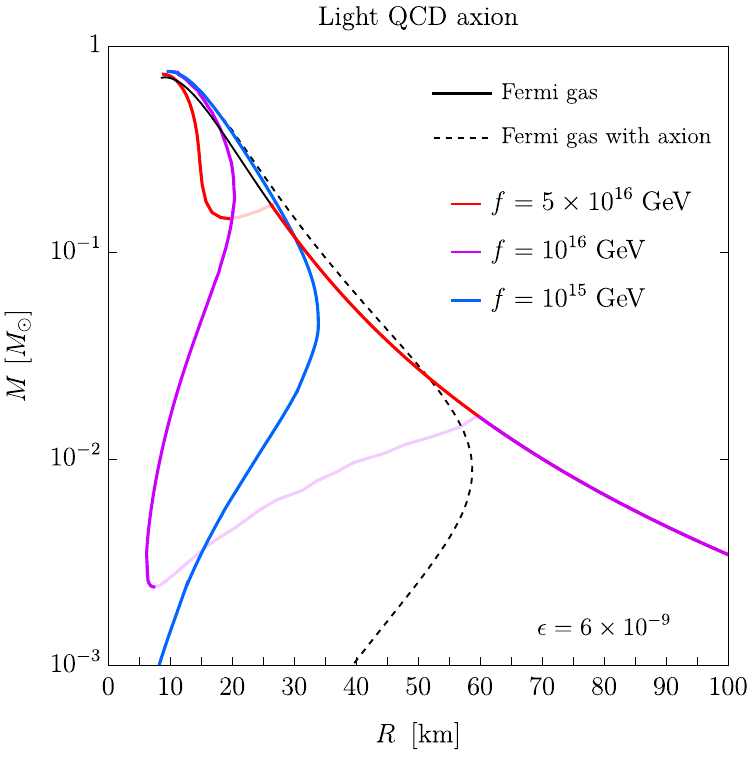} \includegraphics[height=7.5cm]{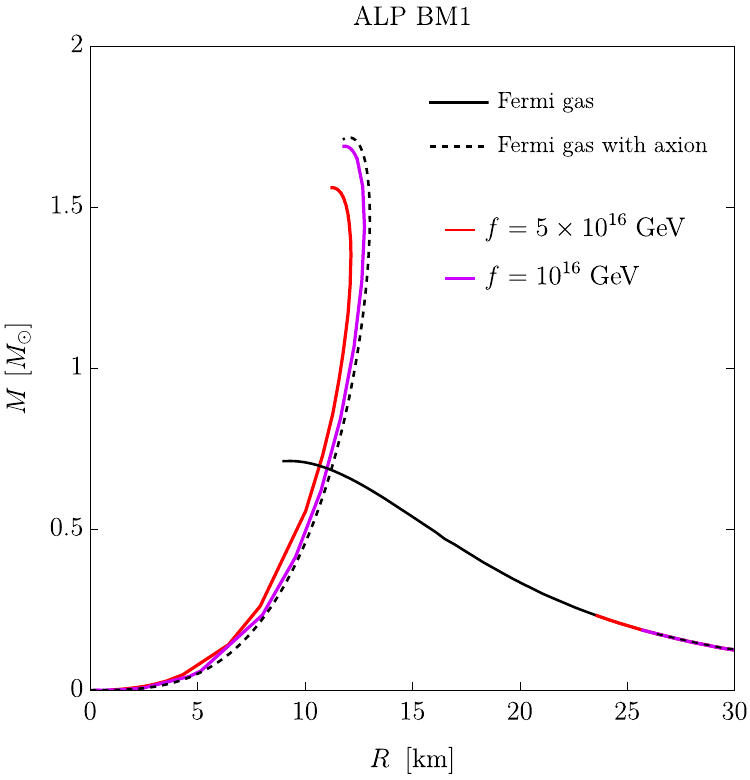}    
    \caption{Left panel: $M$-$R$ curves for a light QCD axions with $\epsilon=6\times10^{-9}$. The free fermi gas without axion (solid black), the negligible gradient limit (dashed black), and including finite gradient effects for $f=\{5\times10^{16},10^{16},10^{15}\} \GeV$ in (solid) red, purple, and blue respectively. The light-colored curves are unstable configurations. Note that the metastable branch for the lowest $f$ is not within the range of the plot. 
    Right panel: $M$-$R$ curves for the ALP benchmark BM1. The free Fermi gas without axion (solid black), the negligible gradient limit (dashed black), and including finite gradient effects for $f=\{5\times10^{16},10^{16}\}\GeV$ in (solid) red and purple, respectively.}
	\label{fig::light_QCD_and_BM}
\end{figure}

Moving beyond the negligible gradient limit, we show the resulting $M$-$R$ curves, found by solving the full coupled system of \Eq{eq::TOV}, in the left panel of \Fig{fig::light_QCD_and_BM}, for different values of $f$. 
Let us note that comparing the analytic estimate in \ref{eq::wavelengthin} with the end of the meta-stable branch, one finds $O(1)$ deviations from finite gradient effects. While expected, this has implications for the robustness of bounds on lighter QCD axions derived from GW observations from the NS merger GW170817 
\cite{Zhang:2021mks}; indeed, assuming similar deviations away from the $\lambda_\phi \ll R$ limit for the NSs in the binary, we find a weakening of the bound at large $f$, precisely where they are relevant, by a factor of a few. Interestingly, we find that the axion halo responsible for the long-range force, which extends much further than the radius of the star (defined by $p_\psi(R)=0$) contributes negligibly to the mass of the system.

While lighter QCD axions lead to a moderate enhancement of the maximal NS mass, they also lead to other striking signatures. The instability of the EOS at intermediate densities, namely between the critical density and NGS density, leads to a gap in the $M$-$R$ curve between the meta-stable branch and the stable branch, clearly visible in \Fig{fig::light_QCD_and_BM}.
As explained in Ref.~\cite{Balkin:2022qer}, the gap moves to smaller radii for higher values of $f$.
Ideally, once the $M$-$R$ plane is sufficiently populated with accurately measured masses and radii of NSs, the observation of such a gap would be a smoking-gun signal for this type of BSM physics, since standard QCD EOSs do not predict such gaps. 
On the other hand, the non-observation of a gap would lead to tight constraints on the parameter space of such models. 
This rationale has been recently followed in Ref.~\cite{Balkin:2022qer}, where the $M$-$R$ distribution of white dwarfs was used, leading to the experimental bound $\epsilon \lesssim 10^{-8}$, stronger than the bounds from the Earth and the Sun \cite{Hook:2017psm}.
We stress that the bounds arising from the existence of a NGS accessible in white dwarfs, and the corresponding gap in radii, are qualitatively very different than the strategy proposed in Ref.~\cite{Hook:2017psm}, which relies on the change of the properties of nuclei, and the corresponding change in X-ray emission, when a (lighter) QCD axion is displaced to $\theta = \pi$ \cite{Ubaldi:2008nf}.
Let us also recall that there are other (weaker) astrophysical and cosmological bounds on $f_a$, see \cite{Raffelt:2006cw} as well as \cite{Lucente:2022vuo,Caloni:2022uya}, which rely on the derivative couplings of axions to the nucleon axial current and the nucleon EDM, which also arise at low energies from \Eq{eq:axionL}.

Another interesting prediction associated with lighter QCD axions is the existence of SBOs (see \Sec{sec::constant_density_objects}), with a range of radii that can potentially span many orders of magnitude, from microscopic to a few km, depending on the value of $f$. At zero temperature, these objects are absolutely stable and cannot decay. We leave the study of the phenomenology of these objects for future work. 

\subsubsection{Axion-like particles}
\label{sec::ALP_model}

Motivated by the interesting phenomenological signatures associated with the existence of a new ground state of matter in lighter versions of the QCD axion, in this section, we wish to explore the possibility that a light scalar has a larger (non-derivative) coupling to nucleons while keeping its potential tunable.
A possible UV completion of such a scenario, based on a $f(\phi)GG$ interaction above the QCD scale, is presented in \App{app::fGG}.

For concreteness, we choose the bounded function $f(\phi)=(1-\cos(\phi/f))/2$, such that the scalar-dependent nucleon mass and the scalar potential are
\begin{align}
\label{eq::m_V_ALP}
m_*(\theta)= m_N\left[1+\frac{g}{2}\left(\cos\theta-1 \right)\right]\,, \;\;\; V(\theta) = -\Lambda^4(\cos\theta-1)\,,
\end{align}
where $\theta \equiv \phi/f$, the dimensionless factor $g$ parametrizes the reduction of the fermion mass when $\theta \neq \theta_0 = 0$, and the scale $\Lambda$ sets the overall scale of the potential.

Following \Eq{eq:mV}, we further identify $n = 2$ and
\begin{equation}
\mphi{2} = \frac{4f^2}{g} \,, \;\;\; 
m_\phi^2 = \frac{\Lambda^4}{f^2} \,, \;\;\; 
F_\phi = f_\phi = f \,,
\label{eq:ALPparam}
\end{equation}
with $\mphi{} \gg F_\phi, f_\phi$ when $g \ll 1$.

As in the previous sections, it is illuminating to first consider the negligible gradient limit. 
At densities above the critical density, implicitly given by 
\begin{equation}
\rho_s(0,\rho_{\text{c}}) = \frac{2 \Lambda^4}{g m_N} \,,
\label{eq:rhocALP}
\end{equation}
the system is minimized at $\theta_{\infty}=\pi$ for $0<g<1$. The (unbounded) case $g\geq1$ will be discussed in \Sec{sec::mstar_unbounded}. For negative $g$, the scalar field will always stay at $\theta_0=0$ since any deviation from the in-vacuum value would result in an increase of the nucleon mass.
As for the (light) QCD axion, scalarization is screened if $g \ll 1$ due to higher-order scalar terms in $m_*$.

We use \Eq{eq::m_V_ALP} as a simple parametrization of any bounded-$m_*$ model that can be mapped to any point in \Fig{fig::ALP_phase_space} according to
\begin{equation}
\label{eq:ALPParam}
m_*(\theta_\infty) = m_N(1-g) \,, \quad
V(\theta_\infty) = 2 \Lambda^4\,.
\end{equation}
For the sake of exposition, let us derive the condition for the existence of the NGS in two different simple limits.
For a non-relativistic Fermi gas, the NGS arises when
\begin{equation}
\label{eq:ALPNGSNR}
\Lambda^4 \lesssim \frac{2 \sqrt{2}}{15 \pi^2} m_N^4 g^{5/2} + O(g^{7/2}) \,, \quad \text{(NR)}
\end{equation}
where we expanded in $g \ll 1$. The corresponding ground state number density is given by
\begin{equation}
\label{eq:rhoALPNGSNR}
\rho^\ngs \sim \left(m_N \Lambda^4 \right)^{3/5} + O(g) \,. \quad \text{(NR)}
\end{equation}
Instead, in the ultra-relativistic limit and expanding around $g = 1$, there exists a NGS for
\begin{equation}
\label{eq:ALPNGSR}
\Lambda^4 \lesssim \frac{m_N^4}{24 \pi^2} + O((1-g)^2) \,, \quad \text{(UR)}
\end{equation}
with the ground state starting at a density
\begin{equation}
\label{eq:rhoALPNGSR}
\rho^\ngs \sim 2 \Lambda^3 \,, \quad \text{(UR)}
\end{equation}
with no $(1-g)$ corrections at leading UR order. Note that although we have chosen $g \ll 1$ for a NR system and $1-g \ll 1$ for an UR one, there certainly exist UR configurations with $g \simeq 0$ and NR ones with $g \simeq 1$.
\\ \\
While the simple ALP model defined by \Eq{eq::m_V_ALP}, in the negligible gradient limit, is a proxy for any model that populates the parameter space $\{m_*(\theta_\infty),V(\theta_\infty)\}$, with the resulting maximal mass and corresponding radii shown in \Fig{fig::ALP_phase_space}, finite gradient effects, on the other hand, are model-dependent; they are sensitive to the particular shape of the scalar potential and the scalar-dependent fermion mass.

To illustrate these finite gradient effects, we consider a benchmark point marked in \Fig{fig::ALP_phase_space} as BM1.
This benchmark is defined by
\begin{equation}
\text{BM1:} \quad g = 0.5 \,, \; \Lambda = 0.075 \GeV \,,
\end{equation}
where we expect to find meta-stable configurations with densities that are at most of the order of the critical density, in this case
\begin{equation}
\rho_c \approx 0.017 \fm^{-3} \approx 0.1\rho_0 \,.
\end{equation}
The absolutely stable configurations are composed purely of the NGS phase, characterized by number and energy densities similar to those of nuclei
\begin{align} 
\label{eq::BM2_NGS_density}
\rho^\ngs \approx 0.17 \fm^{-3} \approx \rho_0\,, \;\;\; \varepsilon^\ngs \simeq m_N \rho^\ngs \approx \varepsilon_0 = 2.5 \times 10^{14} \, \text{g/cm}^3.
\end{align}
Taking gradient effects into account, we find SBOs with their 
(numerically computed)
minimal and maximal radii given by
\begin{align}
\label{eq::rmin_rmax_sbo_BM2}
 R^{\text{\tiny SBO}}_{\text{\tiny min}} &\simeq (25 \, \text{m}) \left( \frac{f}{10^{15} \GeV}\right)\,,
\;\;\;\;
R^{\text{\tiny SBO}}_{\text{\tiny max}} \simeq  (2.3 \km) \left( \frac{f}{10^{15} \GeV}\right)^{1/3}\,.
\end{align}
These values are consistent with the analytical estimates given in \App{app::object_class}.
For objects with radii larger than $  R^{\text{\tiny SBO}}_{\text{\tiny max}}$, gravity becomes relevant and eventually dominant over the scalar force. However, as long as the energy density of the object is approximately constant, the mass and radius are related by \Eq{eq::MR_relation_constant_density}, in this case
\begin{align}
\label{eq::MR_relation_BM2}
M \simeq (5 \times10^{-4} M_{\odot}) \left( \frac{R}{1 \km }\right)^3\,,
\end{align}
which is independent of $f$. 

In the right panel of \Fig{fig::light_QCD_and_BM} we show the $M$-$R$ curves, found by solving the full coupled system of \Eq{eq::TOV}, for $f=\{5\times10^{16},10^{16}\} \GeV$ and in the negligible gradient approximation ($f/\mpl{} \to 0$). At large radii, we find the low-density meta-stable branch corresponding to the free Fermi gas with no scalarization, while at small radii we find the absolutely stable branch, with a large gap (compared to the typical radius) between them.
Note as well that the maximal NS mass with a sourced ALP is much larger than without scalarization.
As shown in the left panel of \Fig{fig::ALP_finite_f_MR_QCD}, for values of $f$ such that scalar gradients are negligible, the enhancement of the maximal mass is even more pronounced for the benchmark denoted by BM2 (dark blue curve), defined by a larger $g = 2/3$ compared to BM1, with the same $\Lambda$.
Indeed,  in the case of large $g$, \ie small $m_*(\theta_\infty)$, this ALP model leads to a large enhancement of the maximal mass of NSs, following \Eq{eq::max_mass_estimate}. 
This is in contrast to the expected reduction in mass due to the softening of the EOS as a result of additional SM degrees of freedom, \eg hyperons or more exotic possibilities such as meson condensation and first-order phase transitions, see \eg \cite{Migdal:1978az,Lattimer:2012nd,Oertel:2016bki}.


\subsection{Unbounded $m_*(\theta)$ solutions}
\label{sec::mstar_unbounded}

We now turn to the analysis of models in which $m_*(\theta)$ is unbounded and vanishes asymptotically, \ie $m_*(\theta = \theta_\infty)=0$, which defines the high-density value of the scalar $\theta_\infty$.
As discussed in \Sec{subsec::micro_EOS_neg_gradient}, $\theta$ approaches $\theta_\infty$ at asymptotically large densities, following the curve in \Eq{eq::theta_inf_case1}.
At such high densities and whenever scalar gradients can be neglected, the maximal mass and corresponding radius only depend on $V(\theta_\infty)$ and are given by \Eq{eq::maxMass_Vinf_scaling}, which recall follows from the ultra-relativistic limit of the EOS. 
In the following, we consider two concrete realizations of such a scenario: linearly- and quadratically-coupled scalar fields, $n = 1$ and $2$ in our classification of \Sec{subsec::chatgpt}.

\subsubsection{Linear coupling to matter}
\label{sec::linear_coup_model}

Let us consider a scalar that couples linearly to nucleons and has a simple quartic potential,
\begin{equation}
m_*(\phi) = m_N\left(1 -  \frac{\phi}{\mphi{}}\right) \ , \quad V(\phi)= \frac{1}{2}m_\phi^2 \phi^2 + \frac{\lambda}{4} \phi^4.
\end{equation}
Following \Eq{eq:mV}, we can easily identify the scales that characterize this model, besides its mass $m_\phi$ and interaction strength with matter $1/\mphi{}$,
\begin{equation}
F_\phi \to \infty \,, \;\;\; f_\phi = \sqrt{\frac{2 m_\phi^2}{\lambda}} \,.
\label{eq:linparam}
\end{equation}
In addition, since the high-density limiting value of the scalar field corresponds to a vanishing $m_*$, it is natural to identify the typical scale of the field as $f \equiv \mphi{}$, thus $\theta \equiv \phi/\mphi{}$.
\\ \\
Let us start by considering the limit $f_\phi \gg \mphi{}$, in which we can neglect the quartic term in the potential for all field excursions, \ie higher-order terms in $\phi$ are irrelevant.
The stellar structure of this model was recently investigated in Ref.~\cite{Gao:2021fyk}. In the following, we re-derive some of the results and pay special attention to the impact of fifth-force bounds.

A linearly-coupled scalar is always sourced at finite density, leading to a fifth force even between dilute and small objects. Despite this fact, as discussed in \Sec{subsec::micro_EOS_neg_gradient} for the unbounded-$m_*$ case, only when densities are of order $\rho_{s,\infty}$ in \Eq{eq::theta_inf_case1} we can expect appreciable effects due to the scalar field being significantly displaced from its in-vacuum value, \ie $\theta \sim \theta_{\infty} = 1$. 
This corresponds to number densities (implicitly) given by
\begin{equation}
\rho_{s}(\theta \sim \theta_{\infty},\rho) \sim \rho_{s,\infty} = \frac{m_\phi^2 \mphi{2}}{m_N} \,,
\label{eq:rholin}
\end{equation}
and assumes finite gradient effects are negligible, \ie
\begin{equation}
R \gg \lambda_\phi \sim \frac{M_\phi}{\sqrt{\rho_{s,\infty}|\partial m_*(\theta)/\partial \theta |}} = m_\phi^{-1},
\label{eq::lin_grad_cond}
\end{equation}
following \Eq{eq::wavelengthin} (taking $\bar \rho_s \sim \rho_{s,\infty}$ and $\Delta \theta \sim 1$).
The parameter space of this model is effectively one-dimensional, since
\begin{align}
\label{eq:LinParam}
m_* \sim m_N\,, \quad V \sim m_\phi^2 M_\phi^2\,,
\end{align}
where recall $m_*(\theta_{\infty}) = 0$, and therefore the dimensionless ratio
\begin{equation}
c \equiv \frac{1}{2} \frac{m_\phi^2 M_\phi^2}{m_N^4} \,.
\label{eq::LinOneDim}
\end{equation}
suffices to describe the phase of the system.
As expected, only for small values of the scalar potential, \ie of $c$, there is a NGS (see the discussion around \Eq{eq:pNGS}), 
\begin{equation}
\label{eq:LinNGS}
c \lesssim 0.015 \equiv c^\ngs \,.
\end{equation}
This transition value between the existence of NGS and the CE region has been found numerically, using the solution of the scalar EOM in the negligible gradient limit, \Eq{eq::microEOS}. We find agreement with the results of Ref.~\cite{Gao:2021fyk}.

\begin{figure}
	\centering
	\includegraphics[width=0.75\textwidth]{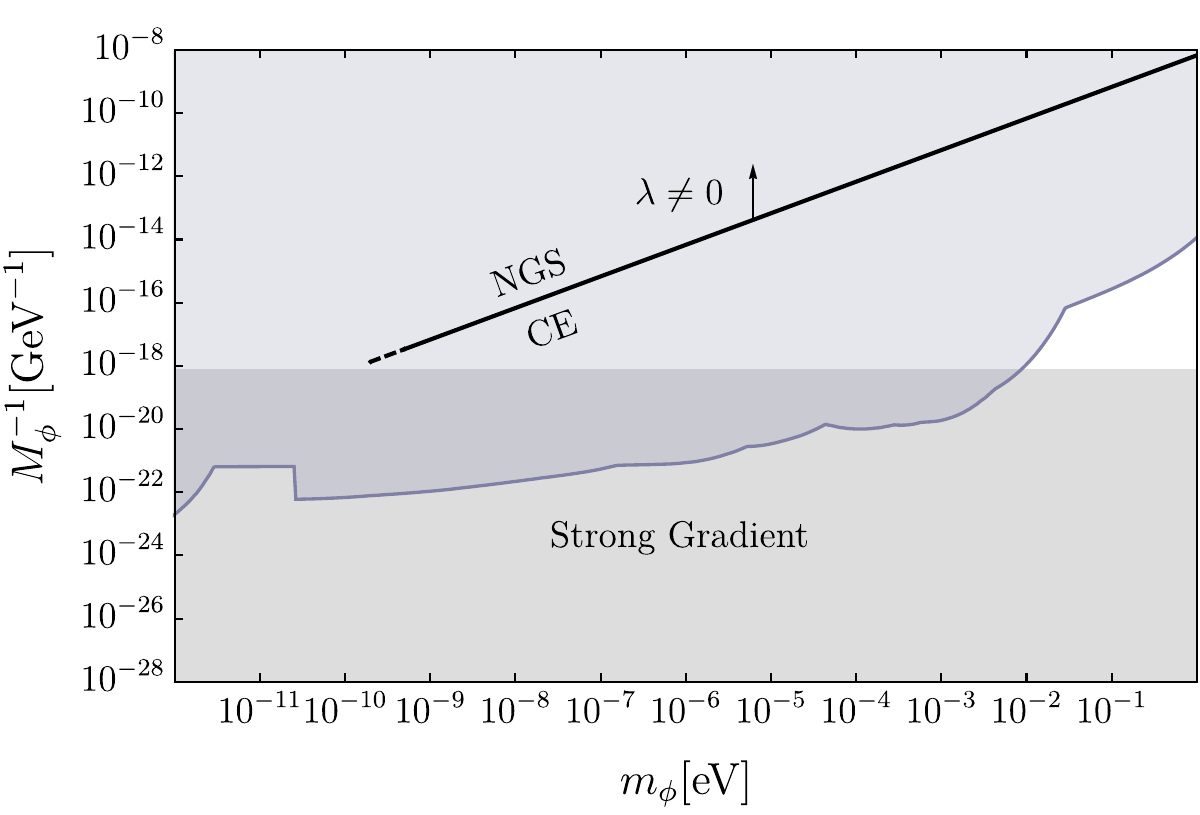}
	\caption{Parameter space of the linearly-coupled scalar model. The black line delimits the transition between the NGS (above) and the CE region (below). The arrow indicates that this transition happens at smaller values of $M_\phi$ for non-negligible $\lambda$. These lines become dashed when neglecting gradient effects in NSs ceases to be a good approximation and ends at the gray-shaded region, which indicates a strong gradient where the scalar field has no effect. In light-blue shaded, we plot fifth-force bounds taken from \cite{OHare:2020wah}, leading to the conclusion that all the interesting parameter space is excluded.}
	\label{fig::linear_plot}
\end{figure}

In \Fig{fig::linear_plot}, we show where the transition between the NGS and the CE region lies in the $\{m_\phi,M_\phi\}$ plane, and compare it with current fifth-force bounds~\cite{OHare:2020wah}, which exclude the blue-shaded region. 
We have dashed the CE-NGS boundary line where gradient effects become relevant, and we have cut it off altogether at the edge of the gray-shaded region, where gradient effects are so strong that the field can no longer be significantly sourced. In this region, the scalar has therefore little to no effect on the configuration of NSs.

In this regard, it is important to take into consideration the size of the scalarized NSs in this scenario. 
Close to the CE-NGS boundary, we find as expected no strong deviations from the standard radii of NSs.
In contrast, deep inside the NGS ($c\ll c^\ngs$), we find \begin{equation}
\label{eq::maxMass_Lin}
 M_{\text{\tiny max}} \sim (0.7\, M_{\odot})\sqrt{\frac{c^\ngs}{c}} \,, \;\;\;
 R(M_{\text{\tiny max}}) \sim (9.3\,\text{km})\sqrt{\frac{c^\ngs}{c}} \,,
\end{equation}
in agreement with \Eq{eq::maxMass_Vinf_scaling}. Given that $\lambda_\phi \sim 1/m_\phi$, see \Eq{eq::lin_grad_cond}, and the parametric estimate for the free Fermi gas radius $R\sim \mpl{}/m_N^2$, this implies that the negligible gradient approximation is valid as long as $M_\phi \ll \sqrt{c^\ngs} \mpl{} \approx \mpl{}/10$.

As can be seen in \Fig{fig::linear_plot}, the parameter space compatible with fifth-force bounds, which extend to $m_\phi \sim \mu\text{m}^{-1}$ for $\mphi{} \sim \mpl{}$, is far from the NGS line. 
Assuming that these bounds can be circumvented, the details of the EOS and stellar structure in this model can be found in Ref.~\cite{Gao:2021fyk}.
However, we expect that a situation where any screening takes place on Earth, such that fifth-force bounds are evaded, yet it does not take place on NSs, is far from generic at the very least.
\\ \\
We conclude this section by considering the effect of the quartic term in the scalar potential. As indicated by the black arrow in \Fig{fig::linear_plot}, a non-negligible $\lambda$, i.e.~$f_\phi \lesssim \mphi{}$, would cause the NGS-CE boundary line to shift up. 
Due to the larger contribution to the energy density and pressure from the potential,
\begin{align}
\label{eq:LinParamQ}
V(\theta_{\infty}) = \frac{1}{2} m_\phi^2 M_\phi^2 \left(1 + \frac{\mphi{2}}{f_\phi^2} \right) \,,
\end{align}
a smaller value of $\mphi{}$ is needed to reduce the fermion mass and reach the NGS.
Such a conclusion is valid only as long as (scalar) densities are above the new density
\begin{equation}
\rho_{s,\infty}^{(\lambda)} =
\rho_{s,\infty} \frac{\mphi{2}}{f_\phi^2} \,,
\label{eq:rholinQ}
\end{equation}
where we have taken $f_\phi \ll \mphi{}$ and $\rho_{s,\infty}$ is given in \Eq{eq:rholin}. For intermediate densities $\rho_{s,\infty}<\rho <\rho_{s,\infty}^{(\lambda)}$, we effectively have a screened system in which $\vev{\phi} \sim f_\phi \ll \mphi{}$.

\subsubsection{Quadratic coupling to matter}
\label{sec::quad_coupling}

Let us consider next a scalar field that couples quadratically to nucleons,
\begin{equation}
\label{eq:mVQuad}
m_*(\phi) = m_N \left(1 -  \frac{\phi^2}{\mphi{2}}\right) \, , \quad V(\phi)= \frac{1}{2}m_\phi^2 \phi^2 + \frac{\lambda}{4} \phi^4 \,.
\end{equation}
Besides the mass $m_\phi$ and the interaction strength with matter set by $1/\mphi{}$, we identify the scales associated with higher-order $\phi$ terms as in the linear model, see \Eq{eq:linparam}. Likewise, the characteristic scale of the field can be conveniently chosen to be $f \equiv \mphi{}$, therefore we define $\theta \equiv \phi/\mphi{}$.%
\footnote{Note that in the $g \gg 1$ limit of the ALP model of \Sec{sec::ALP_model}, fields excursion are small ($\theta \ll1$), thus the ALP model can be mapped to this quadratic model with $m_\phi^2 = \Lambda^4/f^2$, $\lambda = -(\Lambda/f)^4/6$ and $\mphi{2} = 4f^2/g$.}

\begin{figure}[t]
	\centering	\includegraphics[height=7.05cm]{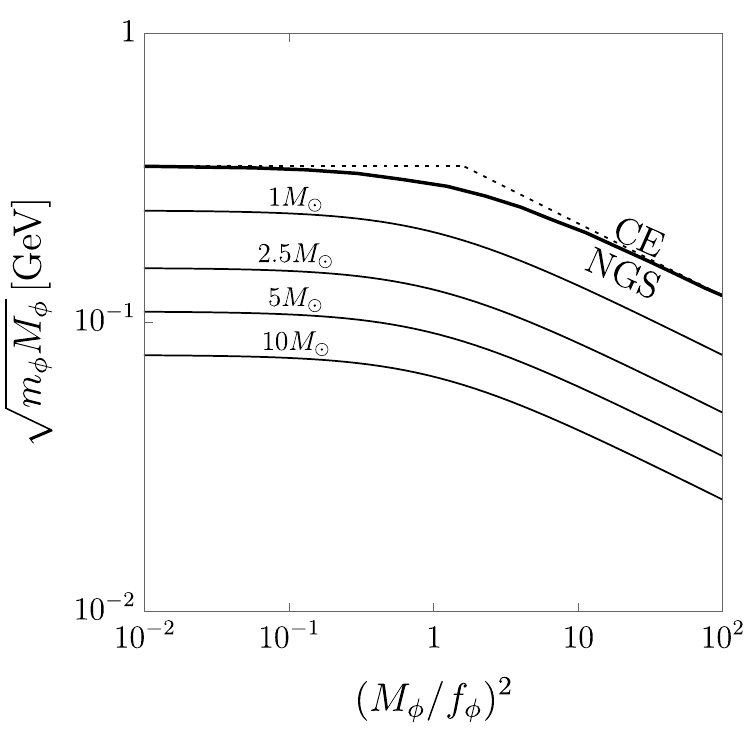} 
	\includegraphics[height=7.1cm]{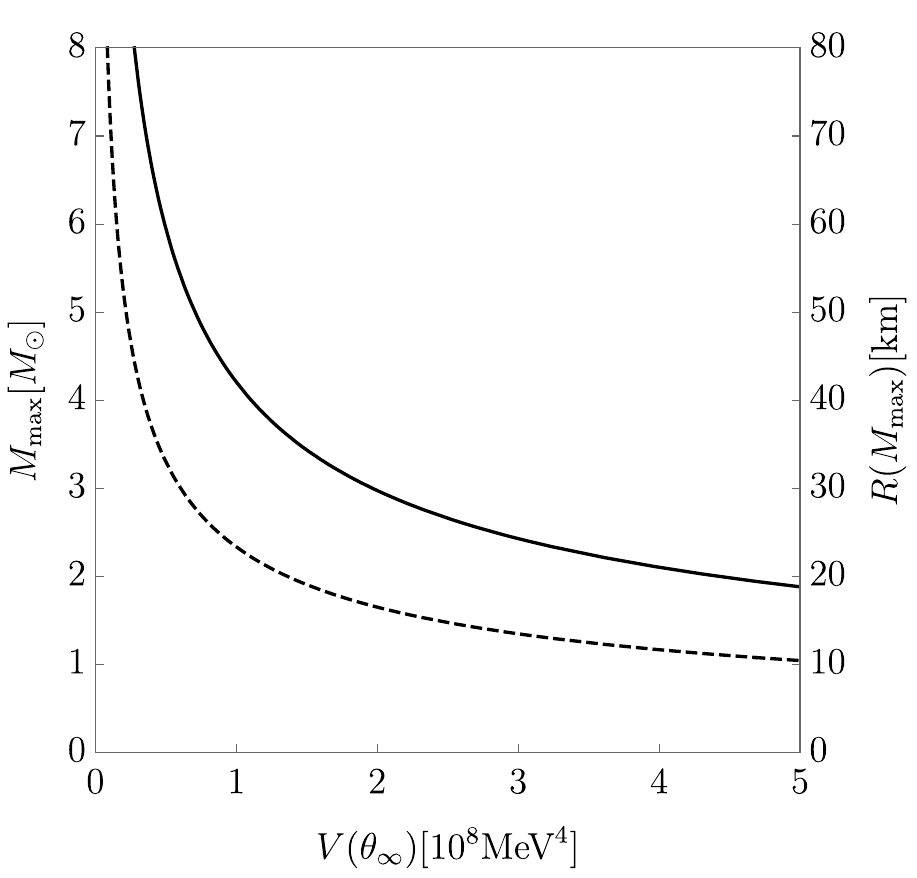}
	\captionof{figure}{Left panel: Parameter space of the quadratic model, with CE above and the NGS below the solid thick line. The dashed line denotes the same NGS-CE boundary but in the $(\mphi{}/f_\phi)^2 \ll 1$ or $(\mphi{}/f_\phi)^2 \gg 1$ limit. Thin-black: contours of maximal mass, for clarity only up to $10 M_\odot$. 
	Right panel: Maximal mass (solid) and corresponding radius (dashed) of NSs as a function of $V(\theta_{\infty})$.}
	\label{fig::parameterquad}
\end{figure}

As in the linear model, we can differentiate between two opposing limits, the mass-dominated regime in which higher-order terms in the scalar potential are irrelevant, $f_\phi^2 \gg \mphi{2}$, and the quartic-dominated regime where instead these control the dynamics, $f_\phi^2 \ll \mphi{2}$.
In each of these regions, the parameter space that determines the phase of the system is effectively one-dimensional,
\begin{align}
c \equiv \frac{1}{2} \frac{m_\phi^2 \mphi{2}}{m_N^4} \,, \quad 
\text{or} \quad 
c_{\lambda} \equiv c \frac{\mphi{2}}{f_\phi^2} = 
\frac{\lambda  \mphi{4}}{4 m_N^4}\,,
\end{align}
for the mass- or quartic-dominated regimes, respectively.
They have a clear physical interpretation as the contributions to the scalar potential in the high-density limit, $V(\theta_{\infty})$, in units of $m_N^4$ (recall $m_*(\theta_{\infty}) = 0$),
\begin{align}
\label{eq:VinftyQuad}
V(\theta_{\infty}) = \frac{1}{2} m_\phi^2 M_\phi^2 \left(1 + \frac{\mphi{2}}{f_\phi^2} \right) 
= m_N^4 \left( c +c_{\lambda} \right)
\,.
\end{align}
As in the linear case, the transition values that separate the CE and NGS regions are found numerically (neglecting gradients). The NGS exists if the dimensionless coefficients satisfy the upper bounds
\begin{align}
\label{eq:cNGSQuad}
c \lesssim 0.0093 \equiv c^\ngs \,, \quad 
\text{and} \quad 
c_{\lambda} \lesssim 0.015 \equiv  c^\ngs_{\lambda} \,,
\end{align}
for $f_\phi^2 \gg \mphi{2}$ and $f_\phi^2 \ll \mphi{2}$, respectively. This is shown as the dashed line in \Fig{fig::parameterquad}.

If both contributions to the potential are of similar size, \ie $f_\phi^2 \sim \mphi{2}$, the parameter space is two-dimensional. The boundary between the CE and NGS region is again found numerically and shown as the solid thick line in \Fig{fig::parameterquad}.

The typical densities associated with the NGS can be estimated according to \Eq{eq::theta_inf_case1}. For the mass- or quartic-dominated regions these are given, respectively, by
\begin{equation}
\rho_{s,\infty} = \frac{m_\phi^2 \mphi{2}}{2 m_N} = m_N^3 c \,, \quad 
\text{or} \quad 
\rho_{s,\infty}^{(\lambda)} = 2 \rho_{s,\infty} \frac{\mphi{2}}{f_\phi^2} = 2 m_N^3 c_\lambda \,.
\label{eq:rhoQuad}
\end{equation}
The scalar field approaches $\theta \to \theta_{\infty}$ asymptotically for densities $\rho \gg \rho_{s,\infty}$ or $\rho \gg \rho_{s,\infty}^{(\lambda)}$,
under the assumption that the field gradient is negligible, \ie $R \gg \lambda_\phi$ with $\lambda_\phi \sim 1/m_\phi$ or $\lambda_\phi \sim f_\phi/(m_\phi \mphi{}) \ll 1/m_\phi$ for a negligible or dominant quartic term, respectively.

Let us also note that in this model, as in the bounded models discussed in \Sec{sec::bounded_models} where the scalars also couple quadratically to matter at leading order in $\phi$, one can identify a critical density where the in-vacuum value of the scalar, $\theta_0 = 0$, becomes unstable, see \Eq{eq:rhoc}.
We find, as expected, that this is given by $\rho_s(0,\rho_c) = \rho_{s,\infty}$.
In addition, in the quartic-dominated regime, \ie $f_\phi^2 \ll \mphi{2}$, we have $\rho_{s,\infty} \ll \rho_{s,\infty}^{(\lambda)}$. At intermediate densities, $\rho_{s,\infty} <\rho< \rho_{s,\infty}^{(\lambda)}$,  the field is destabilized at the origin, $\rho > \rho_c$, thus triggering scalarization, yet $\theta \sim \theta_{\infty}$ is never reached. In such a case, the scalar is screened, and we effectively have $\vev{\phi} \sim f_\phi$ as a limiting value.
\\ \\
Deep inside the NGS region, $\theta \simeq \theta_\infty$, $m_* \ll m_N$, and the maximal star mass and corresponding radius are well approximated by \Eq{eq::maxMass_Vinf_scaling}, with the only relevant dimensionful parameter given by $V(\theta_{\infty})$ in \Eq{eq:VinftyQuad}. 
There are $O(1)$ deviations from these approximations near the NGS-CE boundary, while the deviations are reduced away from the boundary. Numerically, we find \eg that $\theta^\ngs$ is $\sim 10\%$ away from $\theta_{\infty}$ for $c \simeq 10 c^\ngs$, while it is only $\sim 1\%$ away for $c \simeq 100 c^\ngs$, where $ c^\ngs$ are given in \Eq{eq:cNGSQuad} for the mass- and quartic-dominated regimes.
The numerically obtained $M_{\text{\tiny max}}$ and $R(M_{\text{\tiny max}})$ are shown in the right panel of \Fig{fig::parameterquad}.

As opposed to the linearly-coupled scalar, this model leads to large NSs masses (and radii), SBOs, and a gap in the $M$-$R$ curve, without being in tension with current fifth-force bounds (see e.g.~\cite{Brax:2017xho} for a recent analysis on the experimental tests of the forces mediated by quadratically-coupled scalars).
It is worth pointing out that these conclusions rely on the validity of our effective description, \Eq{eq:mVQuad}, at large field excursions, such that higher-order $\phi$ terms in $m_*(\phi)$ do not play any role. This is a non-trivial requirement on the underlying dynamics, especially since we assume our model to be valid towards small nucleon masses.

Finally, we discuss other potentially relevant bounds on this type of scalars bilinearly coupled to nucleons. First, the astrophysical and cosmological bounds that apply to (light) QCD axions (see end of \Sec{sec::lightQCDaxion}), which exhibit as well as quadratic couplings to nuclei, do not generically apply here; these bounds are associated with IR couplings different than those from $m_*(\phi) \bar \psi \psi$, which are predicted from our knowledge of the UV interactions of the scalar above the QCD scale, \Eq{eq:axionL}.
Therefore, without specifying the interactions of the scalar above $\Lambda_{\text{\tiny QCD}}$, such type of bounds do not generically apply. An example of a possible UV completion, based on a coupling to the gluon field strength, is provided in \App{app::fGG}. 
Still, robust, general bounds can potentially be derived given only \Eq{eq:mVQuad}. For instance, as show in Ref.~\cite{Balkin:2022qer}, a reduction of the nucleon mass has a strong impact on the configuration of white dwarfs, whenever these are large and dense enough to trigger scalarization. 
It would be interesting to recast such an analysis to constrain the parameter space of the quadratic model.
For other interesting directions, see e.g.~\cite{Olive:2007aj}, where a bound due to excessive energy loss in supernovae from pair emission of scalars via $N N \to NN\phi\phi$ was estimated to be $\mphi{} \gtrsim 15 \TeV$.
Let us note however that if the scalar is sourced during the supernova, a linear coupling for the $\phi$ excitation is present and suppressed by $\sim 1/\mphi{}$. Accordingly, we estimate the bound from single emission to be $\mphi{} \gtrsim 10^7 \TeV$.


\subsection{Scalar-tensor theories}
\label{sec::scalar-tensor}

Scalar-tensor theories of the type first proposed by Damour and Esposito-Far\`ese \cite{Damour:1993hw} can be equivalently described, as we have done in this paper, via scalar-dependent masses for the (non-interacting) matter fields.
In this section, we re-derive this well-known fact in the context of a free Fermi gas. For a recent discussion, see \cite{DelGrosso:2023trq}.

The action for a conformally coupled scalar field, in the Einstein frame, is given by
\begin{equation}
S = \int d^4x \sqrt{-g} \left[ \frac{\mpl{2}}{16\pi^2} R + \frac{1}{2} g^{\mu \nu} \partial_\mu \phi \partial_\nu \phi - V(\phi) \right] + S_m [\Psi,\tilde{g}_{\mu \nu}] \,, 
\label{eq:Scalartensor}
\end{equation}
with the matter action
\begin{equation}
S_m = \int d^4x \sqrt{-\tilde{g}} \left[ \bar{\Psi} i \tilde{e}^{\mu}_{\,\,\,a} \gamma^a \tilde{D}_\mu \Psi - m \bar{\Psi} \Psi \right] \,, \quad \tilde{g}_{\mu \nu} = A^2(\phi) g_{\mu \nu} \,,
\label{eq:Sm}
\end{equation}
where $\tilde{D}_\mu = \partial_\mu - i \tilde{\omega}_\mu$ is the covariant derivative of a fermion field $\Psi$ in the scalar-dependent metric $\tilde{g}_{\mu \nu}$.
The vielbein and spin connection associated with the matter metric $\tilde{g}_{\mu \nu}$ (a.k.a. Jordan frame) are related to those of the metric $g_{\mu \nu}$ (Einstein frame) as
\begin{equation}
\tilde{e}^{\mu}_{\,\,\,a} = A \, e^{\mu}_{\,\,\,a} \,, \quad \tilde{\omega}_\mu = \omega_\mu + \frac{1}{4} \sigma_{ab} ( e^{\nu b} e^{a}_{\,\,\,\mu} - e^{\nu a} e^{b}_{\,\,\,\mu} ) \partial_\nu \log A \,.
\end{equation}
Noticing in addition that $\sqrt{-\tilde{g}} = A^4 \sqrt{-g}$, one can perform a scalar-dependent field redefinition of the fermion, $\psi \equiv A^{3/2} \Psi$ (known as conformal dressing), such that any dependence of the matter action on $A$ is eliminated except in the mass term, leading to \Eq{eq:lagr} with
\begin{equation}
m_*(\phi) = A(\phi) m \,.
\end{equation}
This shows the equivalence of the two pictures when matter is composed of a free massive field. An equivalent derivation holds if there are more than one species.

At the level of the EOMs and the matter EOS, this equivalence can be seen by noting that the matter source term in the scalar EOM \Eq{eq::gen_KG} is just (working with $\phi = f \theta$)
\begin{equation}
\rho_s \frac{\partial m_*(\phi)}{\partial \phi} = T_\psi \frac{\partial \log A(\phi)}{\partial \phi} \,,
\end{equation}
since $T_\psi = g^{\mu \nu} (T_\psi)_{\mu \nu} = \epsilon_\psi - 3 p_\psi = m_* \rho_s$, which matches the source term of a conformally coupled scalar from \Eq{eq:Sm}.
Furthermore, let us point out the fact that untilded (Einstein frame) quantities being functions of $m_*$ and $\mu$, \eg $p_\psi = p_\psi (m_*(\phi), \mu)$, is consistent with the fact that the energy-momentum tensors in the Einstein and Jordan frames are related by $(T_\psi)_{\mu \nu} = A^2 (\tilde{T}_\Psi)_{\mu \nu}$, thus $p_\psi = A^4 \tilde{p}_\Psi$ with $\tilde{p}_\Psi = \tilde{p}_\Psi(m,\tilde \mu)$ and $\tilde \mu = \mu/A$ (likewise $\tilde{k}_F = \sqrt{\tilde{\mu}^2-m^2} = k_F/A$).
\\ \\
When the matter fields are not free, i.e.~departing from the free Fermi gas limit, encoding all the interactions of a conformally coupled scalar as $\phi$-dependent masses is certainly not enough.
For instance, returning to the case of interest in which the fermion $\psi$ is a nucleon, the introduction of pion-nucleon interactions in the presence of a conformally coupled scalar would require not only scalar-dependent pion masses, $m_{\pi*}^2(\phi) = A^2(\phi) m_\pi^2$, but a scalar-dependent pion decay constant as well, $f_{\pi*}(\phi) = A(\phi) f_\pi$.

This can also be understood by considering the interactions of a conformally coupled scalar above the QCD scale. While classically (neglecting quark masses for simplicity), the QCD action with conformally-dressed fields is a priori independent of $\phi$, the trace anomaly gives rise to the interaction
\begin{equation}
\mathcal{L}_{G\phi} = -\log A(\phi) \frac{\beta(g_s)}{2 g_s} G_{\mu \nu}^2 \,,
\label{eq:anomaly}
\end{equation}
where $G_{\mu \nu}$ is the gluon field strength, $g_s$ the QCD coupling constant and $\beta$ its beta function, $\beta(g_s) = \partial g_s/\partial \log \mu$.
It is this interaction that leads to the $\phi$-dependence of the low-energy parameters after QCD confinement, see \eg \cite{Shifman:1978zn,Kaplan:2000hh,Damour:2010rp}.

This brings us to our final comment, which concerns the radiative stability of a light conformally coupled scalar. The trace anomaly is a manifestation of the fact that scale invariance is not a robust symmetry at the quantum level, with a non-vanishing beta function understood as an explicit breaking of the dilation symmetry. Such a symmetry would naturally ensure a hierarchically small mass for the scalar, if this were identified with a bona-fide dilaton, fixing $A(\phi) = e^{\phi/\mphi{}}$, see e.g.~\cite{Sundrum:2003yt} for a neat discussion.
However, since the symmetry is explicitly broken, one should expect a contribution to the dilaton potential
\begin{equation}
\Delta V(\phi) \sim \frac{M_{\text{\tiny UV}}^4}{16 \pi^2} \frac{\beta}{g_s} 
\frac{\phi}{\mphi{}} \, e^{4\phi/\mphi{}}\,,
\end{equation}
on top of the quartic potential allowed by scale invariance, $V \sim M_{\text{\tiny UV}}^2 \mphi{2} e^{4\phi/\mphi{}}$, with $M_{\text{\tiny UV}}$ the cutoff of the scalar effective field theory.
If instead the function $A(\phi)$ corresponds to a generic conformally coupled scalar, one would expect $\Delta V \sim (M_{\text{\tiny UV}}^2/4\pi)^2 (\beta/g_s) \log A(\phi)$. In any case, the lightness of the scalar, required to yield appreciable effects in stars, is endangered by quantum effects, unless the cutoff is very low or tuning is invoked, see also \App{app::fGG}.
\\ \\
We finish this section with a brief comment on the previous literature. To our knowledge, most works on the configuration of NSs in scalar-tensor theories \Eq{eq:Scalartensor} have focussed in a regime where $\mphi{} \sim \mpl{}$, see \eg \cite{Ramazanoglu:2016kul,Doneva:2016xmf,Brax:2017wcj,Morisaki:2017nit,Staykov:2018hhc} and \cite{Doneva:2022ewd} for a review.
As we have discussed in \Secs{sec::bounded_models}{sec::mstar_unbounded}, in this regime the effects of the scalar gradient cannot be neglected. 
In addition, for the functions $A(\phi)$ and $V(\phi)$ chosen in these works, the NGS of matter is either absent or it has not been identified.
While it would be interesting to reasses these models in light of our new (microscopic) perspective on the scalarized matter EOS, we recall that the NGS is generically present in the so-called strongly scalarized scenario where $\mphi{} \ll \mpl{}$, see \Figs{fig::linear_plot}{fig::parameterquad}.


\section{Conclusions}
\label{sec:conc}

In this work, we have presented a comprehensive and detailed study of the impact of scalarization on the configuration of NSs.
This is a non-trivial back-reaction effect: a dense and large star can source the scalar field, which in turn alters the structure of the star.
We have shown that at leading order the relevant (non-derivative) couplings of the scalar to matter can be encoded as a scalar-dependent nucleon mass.
This allowed us to study in a straightforward way how the EOS of matter, modeled as a free Fermi gas, is affected by scalarization.
In the infinite volume limit, we have shown that the total energy density and pressure of the system receives, beyond the Fermi gas contribution, a contribution from the scalar potential.
It is the interplay between the change in the matter EOS due to a reduction of the nucleon mass and the scalar potential that determines the energetically preferred state of the system.

Our analysis has uncovered what can be considered one of the most salient effects of scalarization: the emergence of a new ground state of (nuclear)
matter at some finite number density and zero pressure. 
The NGS emerges if the change in nucleon mass dominates over the scalar potential, leading to a larger binding energy per nucleon compared to well-separated nucleons.
We have found that the NGS is quite generic and allowed by current constraints by exploring several scenarios beyond the SM with a light scalar: the QCD axion and lighter generalization thereof, generic pseudo-Nambu-Goldstone bosons (which we termed ALPs), and a simple scalar quadratically coupled to nucleons and with a quartic potential.

The phenomenological implications of scalarization, and in particular of the emergence of the NGS, are striking. Because a reduction of the nucleon mass leads to a stiffer EOS, NS masses far beyond the maximal mass predicted by the standard causal bounds can be reached. These stars are also much larger, such that their compactness is approximately the same as that of a free Fermi gas. On the other hand, the contribution of the scalar potential to the energy and pressure softens the EOS. When this effect dominates, we have found stars in a hybrid configuration, where a scalarized core coexists with the rest of the star in the standard phase. NSs with a phase transition within them are lighter yet unusually compact.
In addition, because the standard phase of matter is in fact meta-stable yet (very) long-lived when the scalarized NGS exists, we have found that the $M$-$R$ relation exhibits an instability gap in radii, in which no stars should be found.
In connection to this fact, we have discovered that the new, absolutely stable branch extends down to small self-bound objects in the NGS, held together by the scalar force instead of by gravity. These SBOs have properties, such as mass and compactness, which greatly differ from those of standard stellar remnants.

We have provided analytic, semi-model-independent estimates of key quantities such as critical densities for scalarization, maximal mass of NSs and corresponding radius, as well as minimal and maximal sizes of SBOs. Whenever possible, we have also analytically determined the boundary of the parameter space where the NGS exists, as well as the new ground-state number density of matter, for the different scalar scenarios under consideration. These estimates make the physics transparent, and we have checked all of them against numerical simulations.
In this regard, we paid special attention to finite gradient effects, associated with the non-trivial profile of the scalar in finite volume systems, i.e.~stars in our case. These contributions to the energy density and (anisotropic) pressure are important when the in-medium characteristic wavelength of the scalar field is of the order of the size of the object, and they lead to important departures, especially when the typical scale of the scalar $f$ is close to $\mpl{}$.

We have also made explicit the connection between our analysis and the popular scalar-tensor theories, in the hope that our fresh perspective will contribute to elucidating their full dynamics and whole range of phenomenological implications.
\\ \\
Our results shed new light on the already clear importance of future electromagnetic and gravitational wave observations of compact stellar objects in general and NSs in particular.
Indeed, observatories such as NICER and LIGO have the potential to discover signs, like very heavy NSs and gaps in radii, that could ideally be considered as smoking-gun signals of the scalarized NGS of matter.
Certainly, many degeneracies are at play here, such as astrophysical uncertainties in the expected NS-BH mass gap along with the experimental difficulties in discerning between these two types of compact objects, large experimental errors in the determination of NS radii, or theoretical uncertainties in the EOS of NS matter.
However, many of these constitute in principle a reducible background that could be greatly reduced thanks to the vigorous current and future experimental program in astrophysics.
Besides, while a leap in our theoretical understanding of dense matter is not in foresight, for many of the scalar scenarios we considered, the relevant densities fall within perturbative control.
Indeed, bounds on lighter QCD axions have already been derived from the effects of the NGS on the configuration of the much more dilute white dwarfs \cite{Balkin:2022qer}.
The uncertainty in the EOS is, unfortunately, most pronounced for the case of the QCD axion, where the possibility of scalarization itself is speculative.
Furthermore, let us note that, 
with the exception of the linearly-coupled scalar, 
for all the other models we considered there is wide parameter space open to further experimental exploration. This in turn means these models can be tested with other, complementary probes: pulsar timing, stellar energy loss, long-range forces in binary mergers, or superradiance (see e.g.~\cite{AxionLimits} for an overview of the status of (light) QCD axions).

We conclude with a summary of the directions that we believe deserve further investigation.
An important question that we did not explore concerns the formation of stellar objects in the NGS, for which there are many non-trivial aspects to consider. 
Importantly, its answer is of no consequence to our findings, which concern the long-time, non-dynamical structure of stellar remnants.
A related question concerns the cosmological evolution and phenomenological implications of the SBOs, which can be as small as the Compton wavelength of the scalar.

It would be interesting to extend our analysis to more realistic EOSs. 
We have found, using a free Fermi gas description, that pure neutron matter can be effectively self-bound at high densities due to the scalar dynamics. 
We expect that, by considering a more realistic EOS, this picture does not qualitatively change and asymmetric nuclear matter is self-bound as well. 
From our ongoing work we already have indications that this is the case when including nuclear interactions mediated by pion exchange, even incorporating the effects that scalarization has on the interactions themselves (e.g.~a change in the mass of the pions); we have explicitly checked this for lighter QCD axions. 
Therefore, the emergence of a scalarized ground state of matter seems to be robust in the regime where we retain perturbative control.


\begin{acknowledgments}
The work of JS, KS, SS, and AW has been partially supported by the Collaborative Research Center SFB1258, the Munich Institute for Astro- and Particle Physics \mbox{(MIAPP)}, and by the Excellence Cluster ORIGINS, which is funded by the Deutsche Forschungsgemeinschaft (DFG, German Research Foundation) under Germany's Excellence Strategy -- EXC-2094-390783311. 
The work of JS is supported by the grant RYC-2020-028992-I funded by MCIN/AEI/10.13039/501100011033 and by ``ESF Investing in your future''. JS also acknowledges the support of the Spanish Agencia Estatal de Investigacion through the grant ``IFT Centro de Excelencia Severo Ochoa CEX2020-001007-S''.
The work of SS is additionally supported by the Swiss National Science Foundation under contract 200020-18867. 
The work of RB is supported by grants from the NSF-BSF (No.~2018683), the ISF (No.~482/20), the BSF (No.~2020300), and by the Azrieli Foundation. 
\end{acknowledgments}


\appendix


\section{Dimensional analysis and negligible gradient limit}
\label{app::dim_anal_neg_grad}

It is useful to rewrite the EOMs \Eq{eq::TOV} in terms of dimensionless quantities, which we define as
\begin{align}
&\hat{p} \equiv p/m^4\,, \;\;\; \hat{\varepsilon} \equiv \varepsilon/m^4\,, \;\;\; \hat{r}=r/\alpha\,, \;\;\; \hat{M} = M/(\alpha^3 m^4) \,,
\\
&\hat{V} \equiv V/\Lambda^4\,,\;\;\; \hat{m}_*(\theta) \equiv m_*(\theta)/m\,, \;\;\; \hat\rho_s  \equiv (m\rho_s) /\Lambda^4\,,\nonumber
\end{align}
where $\Lambda^4 \sim m_\phi^2 f^2$ is the typical scale associated with the scalar potential. 
The EOMs are then given by
{
\begin{subequations}
\label{eq::scalar_EOS_dimless}
\begin{align}
\theta''\bigg(1\bigg.&\bigg.-\frac{2c_1 \hat{M}}{ \hat{r} }\bigg)+\frac{2}{\hat{r}}\theta'\left(1-\frac{c_1 \hat{M}}{\hat{r} }- 2\pi c_1 \hat{r}^2 \left(\hat\varepsilon-\hat{p}  \right)\right) = c_3\left( \frac{\partial \hat V}{\partial \theta}+\hat{\rho}_s \frac{\partial \hat{m}_*(\theta)}{\partial \theta}\right),
\label{eq::gen_KG_dimless}\\
\begin{split}
\hat{p}'=&-\frac{c_1\hat{M}\hat\varepsilon}{   \hat{r}^2}\bigg[1+\frac{\hat{p}}{\hat{\varepsilon}} \bigg]\left[1-\frac{2 c_1 \hat{M}}{ \hat{r} }\right]^{-1}\left[1+ \frac{4\pi  \hat{r}^3}{  \hat{M}}\left( \hat{p}+\frac12 c_2  {\theta'}^2\left\{1-\frac{2c_1 \hat{M}}{ \hat{r} }\right\}\right) \right]\\
 &-c_2 c_3 \theta'  \left(\frac{\partial \hat{V}}{\partial \theta}+\hat{\rho}_s \frac{\partial \hat{m}_*}{\partial\theta} \right),
\end{split}
\label{eq::TOV1_dimless}\\
\hat{M}' =&\,\, 4\pi  \hat{r}^2  \left(\hat\varepsilon+\frac12 c_2  {\theta'}^2\left[1-\frac{2 c_1 \hat{M}}{ \hat{r} }\right]\right),\label{eq::TOV2_dimless}
\end{align}
\end{subequations}}
where we identify three relevant dimensionless coefficients
\begin{align}
c_1 \equiv \frac{\alpha^2 m^4}{\mpl{2}}\,, \;\;\; c_2 \equiv \frac{f^2}{\alpha^2 m^4}\,, \;\;\; {c}_3 \equiv \frac{\alpha^2\Lambda^4 }{f^2}\,.
\end{align}
$\alpha$ is an arbitrary length scale, to be chosen at convenience. For instance, the typical size of a gravitationally-bound star is determined by the condition $c_1(\alpha = R)= 1$, which implies $R = \mpl{}/{m^2}$. 
On the other hand, the typical scale associated with the scalar field is given by the condition $c_3\text{Max}\{1,\hat{\rho}_s\}(\alpha=\lambda_{\phi})\sim 1$, which strictly speaking is a locally defined property since $\hat{\rho}_s$ depends on $\hat{r}$. This typical scale of the scalar field, which changes with density, is what we refer to as the wavelength of the field. 


\subsection*{The negligible gradient approximation and in-medium wavelength }
\label{app::negligible_grad}

To understand the negligible gradient approximation, let us focus on \Eq{eq::gen_KG_dimless},
\begin{equation}
\theta'' + \frac{2}{\hat{r}} \theta' = c_3\left(\frac{\partial \hat{V}}{\partial \theta}+\hat{\rho}_s \frac{\partial \hat{m}_*}{\partial\theta} \right) \,,
\label{eq::scalar_EOM_approx_nograv}
\end{equation}
where we neglected the $O(1)$ deformation of the scalar derivatives due to gravity, as they do not play a significant role in the following discussion. Over a region of size $\hat{r}_2-\hat{r}_1\equiv\Delta \hat{r}$, around the mean position $\frac12 (\hat{r}_2+\hat{r}_1) \equiv \hat{r}_{c}$ where $\theta$ changes by $\Delta \theta$, the LHS scales as $\sim \frac{\Delta \theta}{(\Delta \hat{r})^2}$. This is true both in the case where the transition occurs when $ \hat{r}_{c} \gg \Delta \hat{r}$, in which case the $\theta'$ term is $O(\Delta \hat{r}/\hat{r}_{c})$ suppressed w.r.t~to the $\theta''$ term, or when the transition happens for $\hat{r}_{c} \sim \Delta \hat{r}$, in which case the $\theta'$ and $\theta''$ term scale the same.

Consider then the scalar profile $\theta(r)$ derived from the EOS \Eq{eq::microEOS}, where $\rho_s = \rho_s(r)$ as follows from the solution of \Eq{eq::TOV_w_microEOS}. By construction this ensures that the RHS of \Eq{eq::scalar_EOM_approx_nograv} vanishes. This is a good approximation to the full coupled system of EOMs if the corrections due to the scalar field derivatives $\theta'(r)$ and $\theta''(r)$ in \Eq{eq::scalar_EOS_dimless} can be considered small, in which case the EOMs reduce indeed to \Eq{eq::microEOS} and \Eq{eq::TOV_w_microEOS}.
Let us assess the validity of this approximation by separating the discussion into two qualitatively different cases depending on the behavior
of $\theta(r)$, (1) $\theta$ varies continuously in a finite region and (2) $\theta$ is discontinuous, i.e.~jumps from one value to another, forming a so-called bubble wall. In both of these regions, we argue that while $\theta'$ and $\theta''$ do not vanish, $\theta(r)$ can nonetheless be considered a good approximate solution overall under certain conditions. We shall derive an upper bound on $f$ which ensures these conditions are satisfied, with the strongest bound coming from the condition for the formation of the bubble wall.
\\ \\
We start with the case where $\theta(r)$ is continuous. This is typical in linearly coupled models ($n = 1$ in \Eq{eq:mV}, see \Sec{sec::linear_coup_model}), at least at small enough densities.
While it is easier to characterize the gradient corrections in this region, as we show below it typically provides a weaker bound on $f$ when both (1) and (2) behaviors of $\theta$ can happen within the same scalar theory. Consider a region where $\theta(r)$ 
undergoes an $O(1)$ change in its value from $\theta_0$ to $\theta_{\infty}$ such that $\bar{\theta}\equiv (\theta_{\infty}+\theta_{0})/2 \sim \Delta\theta \equiv \theta_{\infty}-\theta_{0} = O(1)$, within a region of size $\Delta\hat{r}$.
The LHS of \Eq{eq::scalar_EOM_approx_nograv} scales like $\Delta \theta/\Delta\hat{r}^2$, as previously explained.
Since the RHS vanishes at leading order by construction, it is sensible to Taylor expand it and evaluate it at $\bar\theta$. By demanding that the change in $\theta$ takes place within the confines of a star, \ie $\Delta \hat{r} \lesssim 1$ in units $\alpha = R$, and that deviations from the assumed profile are at most of order $\Delta \theta$, we arrive at the condition
\begin{align}
R^2 \gg \frac{1}{m_\phi^2(\bar{\theta})} \;\;\; \text{where}\;\;\; m_\phi^2(\bar{\theta}) f^2 \equiv \frac{\partial}{\partial \theta} \left(\frac{\partial V(\theta)}{\partial \theta}+{\rho}_s \frac{\partial {m}_*}{\partial\theta} \right)\biggr|_{\theta=\bar\theta}\,.
\label{eq:mphieff}
\end{align}
This condition can also be interpreted as an energy requirement, identifying ${f^2}/{R^2}$ as the gradient energy density associated with the smooth change of the scalar field, and $m_\phi^2(\bar{\theta}) f^2$ at the gain in effective potential energy.
Note that these types of continuous transitions also occur in models where the $\theta_{\infty}$ phase is ultra-relativistic, which is the common case in the models in \Sec{sec::mstar_unbounded}, as well as in part of the parameter space of the models in \Sec{sec::bounded_models}.
\\ \\
A different condition associated with the negligible gradient approximation, typically leading to a stronger upper bound on $f$, corresponds to the case in which $\theta(r)$ exhibits a discontinuity. This happens for instance in quadratically coupled models ($n = 2$ in \Eq{eq:mV}) just above the critical density for scalarization.
The gradient energy density associated with such a jump in $\theta$ is naively infinite, since $\theta'$ and $\theta''$ are singular. Clearly, this is not a sensible result, and indeed the bubble wall is not infinitely thin but it has a finite size, determined by the in-medium wavelength of the scalar field, $\lambda_\phi$.
Since in the transition region (\ie inside the wall), $\theta(r)$ solving \Eq{eq::microEOS} is not a good approximation, we return to \Eq{eq::scalar_EOM_approx_nograv} and set the units to $\alpha = \lambda_\phi$. 
We then find that at low densities, the wavelength can be estimated as
\begin{align}
\label{eq::wavelengthout}
\lambda_{\phi}^{\text{\tiny low}}  \equiv \frac{\sqrt{\Delta \theta} f}{\sqrt{(\partial V/\partial \theta)|_{\theta = \bar\theta } }} \,,
\end{align}
where $\Delta \theta \equiv \theta_{\infty}-\theta_{0}$ is the jump in $\theta$ and $\bar{\theta}\equiv (\theta_{\infty}+\theta_{0})/2$, and as before we generically consider $\Delta \theta \sim \bar{\theta} = O(1)$.
By using this definition, as opposed to the vacuum Compton wavelength defined as $m_\phi^{-1} \equiv f/\sqrt{(\partial^2 V/\partial \theta^2)|_{\theta=\theta_0}}$, we avoid potentially misidentifying the relevant scaling of the RHS of \Eq{eq::scalar_EOM_approx_nograv},  which can be dominated by higher order terms in the potential once $\theta$ is sufficiently far away from $\theta_0$. This can occur for scalar potentials that feature more than one scale, like in the quadratic coupling model discussed in \Sec{sec::quad_coupling}, where the potential can be dominated by the quartic term. This effect, which admittedly  requires some fine-tuning in the potential, is nonetheless captured by the definition in \Eq{eq::wavelengthout}. In natural potentials defined by a single scale, i.e. $f_\phi \sim f$, $\lambda_{\phi}^{\text{\tiny low}}$ reduces to $m_\phi^{-1}$ as expected, like in the models discussed in \Sec{sec::bounded_models}.

We emphasize however that $\lambda_{\phi}^{\text{\tiny low}}$ does not actually play an important 
role in determining the upper bound on $f$ from the requirement of negligible gradient energy, but rather the in-medium or high-density effective wavelength, identified as
\begin{equation}
\lambda_{\phi} \equiv \frac{\sqrt{\Delta \theta} f}{\sqrt{\bar{\rho}_s|\partial m_*/\partial \theta||_{\theta=\bar{\theta}}}}\,,
\label{eq::wavelengthin}
\end{equation}
where $\bar{\rho}_s$ is the typical scalar density at the internal edge of the transition region, \ie roughly the scalar density at the lowest pressure of the internal phase.
The fact that $\lambda_{\phi}$, rather than $\lambda_{\phi}^{\text{\tiny low}}$, is what determines the relevant size of the scalar bubble stems from the requirement that, for the scalar field to be significantly sourced, a sufficiently large scalar density is needed, in particular $m \rho_s \gtrsim m_\phi^2 f^2 (\mphi{}/f)^n$, following the discussion after \Eq{eq::microEOS} and using \Eq{eq:mV}.
Still, $\lambda_{\phi}^{\text{\tiny low}}$ is interesting in cases where $\lambda_{\phi}^{\text{\tiny low}} \gtrsim R$ (while $\lambda_\phi < \lambda_{\phi}^{\text{\tiny low}}$). In this case, the scalar profile extends to the region outside the star, resulting in a scalar halo and potentially interesting effects like long-range forces between stars \cite{Hook:2017psm}.

Having determined $\lambda_\phi$, note that the gradient energy density associated with a bubble wall is typically larger than that of a continuous scalar profile discussed above, given that the former scales as $f^2/\lambda^2_\phi$, thus enhanced w.r.t.~the latter by $(R/\lambda_\phi)^2$. Nevertheless, the condition
\begin{align}
\label{eq::neg_grad_limit}
\lambda_\phi \ll R\,,
\end{align}
is the correct one to ensure first that the scalar can be sourced when $\theta(r)$ solving \Eq{eq::microEOS} is discontinuous, and second that its contribution to the TOV equations \Eq{eq::TOV1_dimless} and \Eq{eq::TOV2_dimless} is subleading. For $\lambda_\phi \gg R$, the formation of the bubble is not energetically favorable, leading to the trivial $\theta=\theta_0$ solution. This regime can be understood as a decoupling limit $f/\mpl{}\gg 1$, given the scaling with $f$ of \Eq{eq::wavelengthin} and that $R \sim \mpl{}/m^2$. For $\lambda_\phi \sim R$, the effects of the gradient energy are no longer negligible and the full coupled system of \Eq{eq::scalar_EOS_dimless} must be solved (also when $1/m_\phi(\bar \theta) \sim R$ in \Eq{eq:mphieff}). Therefore, only when \Eq{eq::neg_grad_limit} holds the bubble can be formed, and its contribution to the total energy (mass) of the star be safely neglected, given that it is localized to a thin region much smaller than the total size of the star.



\section{Analytic discussion of constant density objects}
\label{app::object_class}

In this appendix, we derive the main results presented in \Sec{sec::constant_density_objects}. In models that allow a NGS, the total core pressure for all configurations on the stable branch balances two distinct contributions,
\begin{align}
p(0) \equiv p_0 = \Delta p_{\text{\tiny grav.}}+\Delta p_{\text{\tiny grad.}}\,.
\end{align}
with
\begin{align}
\Delta p_{\text{\tiny grad.}} &=   -\int^R_{0}\mathrm{d}r\, \theta' \left(\frac{\partial V}{\partial \theta}+\rho_s\frac{\partial m_*}{\partial \theta}\right)\,,
\\
\Delta p_{\text{\tiny grav.}} &\simeq   -\int^R_{0}\mathrm{d}r\, \frac{M(r)\varepsilon(r)}{ \mpl{2} r^2}\,,
\end{align}
where we took for simplicity the Newtonian limit of the TOV equations, which, as discussed below, suffices for the estimations in this section. 
\\ \\
Self-bound objects (SBOs) are such that $\Delta p_{\text{\tiny grav.}}\ll  \Delta p_{\text{\tiny grad.}}$, they are composed of matter in the NGS and are held together by the gradient pressure of the scalar field.
The opposite limit, $\Delta p_{\text{\tiny grav.}}\gg \Delta p_{\text{\tiny grad.}}$, corresponds to stars.
The smallest possible configurations on the stable branch are of size $R \sim \lambda_\phi$, and both gravitational and gradient pressures are spread across the object, thus $\Delta p_{\text{\tiny grav.}}$ and $\Delta p_{\text{\tiny grad.}}$ must be calculated by summing up the contributions from $r=R$ until the core. For a given $f$, these smallest configurations can either be self- or gravitationally-bound.
\\ \\
In order to analytically characterize, up to $O(1)$ factors, these objects in different regimes, we make the following simplifying assumptions. First, we consider large systems, $R \gg \lambda_\phi$, in which $p'_{\text{\tiny grad.}}$ is localized at the boundary of the object where the transition occurs. Second, we assume a simple linear energy density profile from the core $\varepsilon(0) \equiv x_0 \varepsilon_\ngs$ to the edge of the object $\varepsilon(R) \equiv x_R \varepsilon_\ngs$,
\begin{align}
\varepsilon(r )/\varepsilon_\ngs = (x_0-x_R)(1-r/R)+x_R\,.
\end{align}
Next, we calculate the gravitational pressure using the Newtonian limit of the TOV equations, neglecting $O(1)$ general relativistic corrections as well as the contribution of the localized gradient energy to the energy density. Lastly, for simplicity, we assume the NGS phase is non-relativistic.
This approximation is typically applicable when $m_*(\theta)$ is bounded, in which case the NGS can be anything between non- and ultra-relativistic. For an ultra-relativistic NGS phase, a similar derivation is straightforward. When $m_*(\theta)$ is unbounded, the NGS is typically ultra-relativistic and has a fixed $\rho_s$.  Also in this case, a similar derivation is straightforward. 

Under these assumptions, we can approximate the internal pressure as
\begin{align}
p_0 &\simeq \underbrace{\frac{ R^2 \varepsilon_\ngs^2 x_0^2 }{\mpl{2}}}_{\Delta p_{\text{\tiny grav.}}}
+ \underbrace{\frac{ f\sqrt{ \varepsilon_\ngs x_R }}{R} \sqrt{\frac{\dm}{1-\dm}}}_{\Delta p_{\text{\tiny grad.}}} 
\simeq \frac{\varepsilon_\ngs^{5/3} (x_0^{5/3}-1)}{m_*^{8/3}}\,,
\label{eq::pressure_grav_grad}
\end{align}
where $\Delta p_{\text{\tiny grad.}}$ is calculated over a small transition region from $R$ to $R+\lambda_\phi$,  approximating the scalar profile as a linear transition, \ie $\theta' \sim 1/\lambda_\phi = \text{const.}$, 
and taking the leading order term in the $\lambda_{\phi}/R\ll 1$ expansion. We also approximated $\frac{\partial m_*}{\partial \theta} \sim m\,\dm$.
For simplicity, we assume in the following that $\dm \ll1$. This assumption can be relaxed by rescaling $\dm \to \dm/(1-\dm)$ in all the expressions below.
Finally, the last equality in \Eq{eq::pressure_grav_grad} is a self-consistency condition due to the assumed non-relativistic EOS at the core, \ie $p_0 = p(\varepsilon(0))$.


\subsection*{Self-bound objects: $\Delta p_{\text{\tiny grav.}}\ll \Delta p_{\text{\tiny grad.}}$}

An interesting prediction for finite $f$ is the existence of SBOs, for which gravity does not a play role, and therefore their pressure, energy density, and scalar field profiles can be computed by solving the $\mpl{}\to\infty$ limit of the coupled TOV equations, \Eq{eq::TOV_selfbound_txt}.
In the $R \gg \lambda_{\phi}$ limit, they are well-described by constant profiles.
They are held together by the gradient pressure exerted at the edge of the object, where the transition occurs in the form of a scalar bubble wall of size $\lambda_\phi$ that ``traps'' the matter inside and prevents it from expanding. 

We apply the simple model of \Eq{eq::pressure_grav_grad} by taking $x_0 = x_R \equiv x_{\text{\tiny SBO}}\gtrsim1$, which describes a constant energy density system.
This allows us to write the SBO radius as a function of its core energy density (or equivalently, in this approximation, its core number density)
\begin{align}
R_{\text{\tiny SBO}}(x_{\text{\tiny SBO}}) \simeq \frac{m_*^{8/3} \dm^{1/2}f}{ \varepsilon^{7/6}_\ngs}  \frac{x_{\text{\tiny SBO}}^{1/2}}{x_{\text{\tiny SBO}}^{5/3}-1}
\label{eq::radius_density_SBO_new}\,,
\end{align}
where $\Delta p_{\text{\tiny grav.}}$ was neglected.
From \Eq{eq::radius_density_SBO_new}, it is clear that SBOs become smaller (larger) as the central density, \ie $x_{\text{\tiny SBO}}$, increases (decreases).
Their total mass is given by
\begin{align}
M_{\text{\tiny SBO}}(x_{\text{\tiny SBO}}) = \frac{4 \pi}{3} x_{\text{\tiny SBO}} \varepsilon_\ngs R^3_{\text{\tiny SBO}}(x_{\text{\tiny SBO}}) \simeq \frac{4 \pi}{3} \frac{f^3 \dm^{3/2} m_*^{8}}{\varepsilon^{5/2}_\ngs}\frac{ x_{\text{\tiny SBO}}^{5/2} }{ (x_{\text{\tiny SBO}}^{5/3}-1)^3}\,.
\end{align}
For finite $f$, SBOs are bounded in size both from above and from below. The smallest object possible would have $R_{\text{\tiny SBO}}^{\text{\tiny min}} \sim \lambda_\phi$, for which the approximation $R\gg \lambda_\phi$ breaks down. This also implicitly defines the maximal possible density of the self-bound object. 
For very low densities the object becomes large, reaching the point where $\Delta p_{\text{\tiny grav.}}$ can no longer be neglected. We then define the maximal SBO radius at the equilibrium point  $\Delta p_{\text{\tiny grav.}}\simeq \Delta p_{\text{\tiny grad.}}$. Using \Eq{eq::pressure_grav_grad}, we can analytically estimate it in two limiting cases
\begin{align}
R_{\text{\tiny SBO}}^{\text{\tiny max}} 
\sim\begin{cases}
\left( \frac{f \mpl{2} \dm^{1/2}  }{\varepsilon^{3/2}_\ngs} \right)^{1/3}\;\;\;\;\;\;\;\;\;\; & 
\dm^{1/2} \left( \frac{m_*^4}{\varepsilon_\ngs} \right) \left(\frac{f}{\mpl{}}\right) \ll 1
\\
\left(\frac{\mpl{7}}{\dm^{1/2} m_*^{12} f}\right)^{1/6} & 
\dm^{1/2} \left( \frac{m_*^4}{\varepsilon_\ngs} \right) \left(\frac{f}{\mpl{}}\right) \gg 1
\end{cases}\,.
\end{align}
The first case corresponds to $x_{\text{\tiny SBO}}\simeq 1$, in which the equilibrium happens when the SBO is close to its ground state density. The second case corresponds to $x_{\text{\tiny SBO}}\gg 1$, in which the equilibrium happens when the SBO is much denser than its ground state density, and thus the maximal radius is independent of the NGS properties.


\subsection*{Constant energy density gravitationally-bound objects: $\Delta p_{\text{\tiny grav.}}\gg \Delta p_{\text{\tiny grad.}}$}
As the core pressure increases, gravity pressure becomes the dominant component. 
We return to our simple model of \Eq{eq::pressure_grav_grad}, and note that, in the absence of a sizeable gradient pressure, vanishing Fermi pressure at $r=R$ requires us to set $x_R \simeq 1$, while we have $x_0 \gtrsim1$, finding
\begin{align}
R(x_0)
\sim \left(\frac{\mpl{2} (x_0^{5/3}-1)}{ m_*^{8/3} \varepsilon^{1/3}_\ngs x^2_0}\right)^{1/2} \,.
\label{eq:Rgrav}
\end{align}
Around $x_0\simeq 1$, the radius increases with increasing core energy density, in contrast to the SBOs which exhibited the opposite behavior. Still, in the limit, $\dm^{1/2}(m_*^4/\varepsilon_\ngs)(f/\mpl{}) \ll 1$, these smallest gravitationally-bound systems can also be approximated as of constant energy density since although their pressure drops away from the core, it is sufficiently low that the EOS is always close to $\varepsilon \simeq \varepsilon_\ngs$.
At large enough core pressures, the constant energy density approximation breaks down, and any further increase in core pressure leads to a decrease in radius, which is the typical behavior for gravitationally-bound objects described by a Fermi gas; 
indeed, the radius \Eq{eq:Rgrav} decreases with increasing core energy density for $x_0 \gg 1$.

Finally, recall that for $\dm^{1/2}(m_*^4/\varepsilon_\ngs)(f/\mpl{}) \gg 1$, the most massive and largest SBOs have $\varepsilon \gg \varepsilon_\ngs$ (i.e.~$x_{\text{\tiny SBO}} \gg 1$).
Therefore, any increase in pressure would lead to a swift breakdown of the constant energy approximation as follows from the EOS, and a return to the typical radius decrease with increasing core pressure.
As a result, the maximal mass of the gravitationally-bound stars in the NGB coincides with that of the SBOs in this limit. 

To conclude, we can approximate the maximal star radius, up to $O(1)$ factors, as
\begin{align}
\label{eq::RGBO_max_estimate}
R^{\text{\tiny max}}
\sim \begin{cases}
\left(\frac{\mpl{2}}{m_*^{8/3} \varepsilon^{1/3}_\ngs}\right)^{1/2} = \left( \frac{\dm^{1/2} m_*^4 f}{\varepsilon_\ngs \mpl{}}\right)^{-1/3}R_{\text{\tiny SBO}}^{\text{\tiny max}}  \;\;\;\;\;\;\;\;\;\; & 
\dm^{1/2} \left( \frac{m_*^4}{\varepsilon_\ngs} \right) \left(\frac{f}{\mpl{}}\right) \ll 1
\\
\left(\frac{\mpl{7}}{\dm^{1/2} m_*^{12} f}\right)^{1/6} = R_{\text{\tiny SBO}}^{\text{\tiny max}}  &  
\dm^{1/2} \left( \frac{m_*^4}{\varepsilon_\ngs} \right) \left(\frac{f}{\mpl{}}\right) \gg 1
\end{cases}\,,
\end{align}
which are the analytic estimates  used in the main text in \Eq{eq::size_of_constant_objects}.


\subsection*{Comparison with numerical results of BM1 ALP model}

\begin{figure}[t]
     \centering    
     \includegraphics[height=7.5cm]{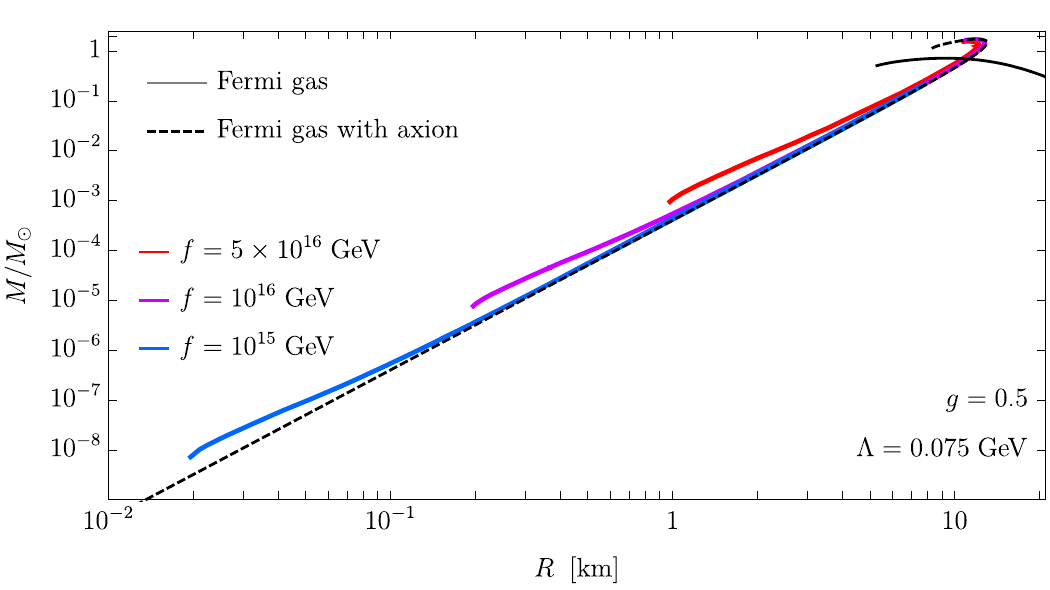}   
     \caption{The low-mass and small-radius region of the $M$-$R$ curves of the benchmark BM1 in the ALP model of \Sec{sec::ALP_model}. The free Fermi gas without axion (solid black), the negligible gradient limit (dashed black), and including finite gradient effects for $f = \{5\times10^{16},10^{16},10^{15}\} \GeV$ in red, purple, and blue respectively.}
     \label{fig::BM_log}
\end{figure}

Let us check our analytical estimates with the numerical results we presented for the BM1 benchmark in \Sec{sec::ALP_model}. For this purpose, it is useful to zoom in on the right panel of \Fig{fig::light_QCD_and_BM} at low masses and small radii, see \Fig{fig::BM_log}. 
Our numerical results agree with our estimates for $ R^{\text{\tiny SBO}}_{\text{\tiny min}}$ given in \Eq{eq::rmin_rmax_sbo_BM2} and are indeed well-described at low pressures by the curve defined by \Eq{eq::MR_relation_BM2} independently of $f$.
The visible deviations from the line at low radii is a finite gradient effect explained by our modeling of the SBOs described in this appendix.
The smallest SBOs can have a central number and energy densities which can be a few times larger than $\rho_\ngs$ and $\varepsilon_\ngs$, respectively. 
Therefore, the energy density can be larger than $\varepsilon_\ngs$, leading to configurations that lie above the curve defined by \Eq{eq::MR_relation_BM2}. 
For the smallest objects, with $R \simeq R^{\text{\tiny SBO}}_{\text{\tiny min}}$, the size of the transition region becomes comparable to the size of the object, and the assumptions of constant pressure and number density, on which our description depends on, are no longer valid. 
Note that the dashed line describing the $M$-$R$ curve in the negligible gradient limit ($f/\mpl{} \to 0$) describes gravitationally-bound constant-density objects since gravity is the only force in this limit. However, as long as the object is dilute enough, \ie $\rho \simeq \rho_\ngs$, it has constant energy density $\varepsilon_\ngs$ and \Eq{eq::MR_relation_BM2} is valid, regardless of whether it is a self- or gravitationally-bound object. 
Note also that for the most massive NGS stars, the effects of the gradient pressure at the edge of the star is increasingly negligible, making the properties of the gravitationally-bound stars essentially $f$-independent, clearly visible in the right panel of \Fig{fig::light_QCD_and_BM}.
\begin{figure}
     \centering
     \includegraphics[width=0.45\textwidth]{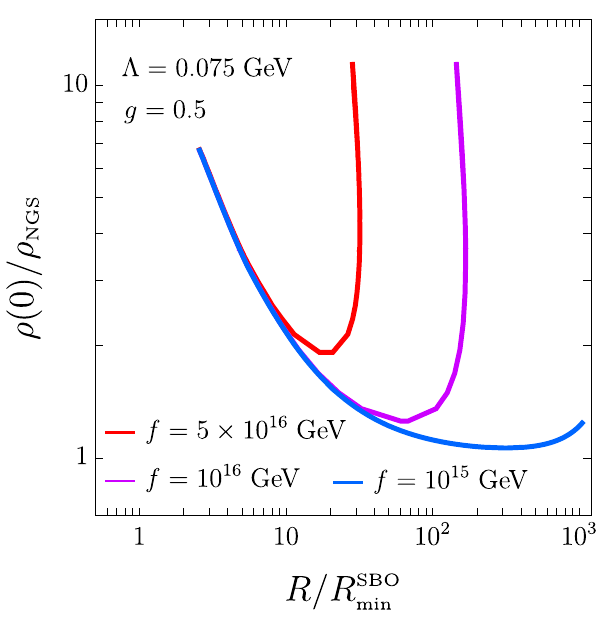}
     \includegraphics[width=0.45\textwidth]{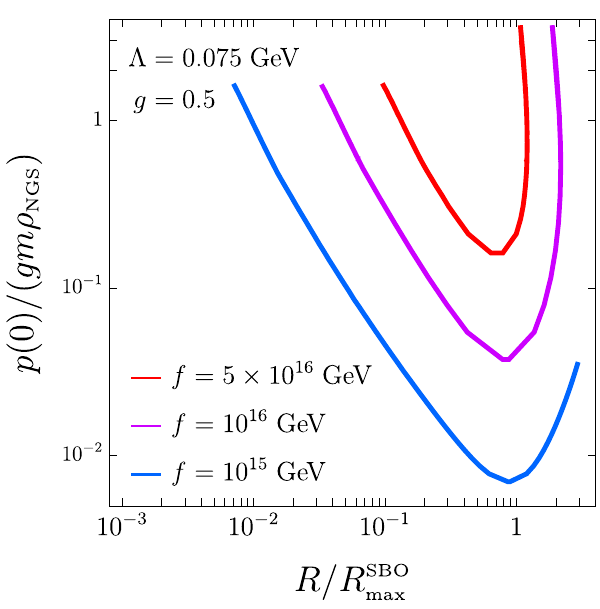}
     \caption{Left panel: The NGS configurations in the $\{\rho(0),R\}$ plane, where $\rho(0)$ is the central number density. $\rho(0)$ is shown in units of the ($f$-independent) NGS number density given in \Eq{eq::BM2_NGS_density}, while the radius is given in units of the ($f$-dependent) minimal radius expected for the SBOs, given in \Eq{eq::rmin_rmax_sbo_BM2} . The numerical results agree with the analytical estimates of the maximal density and minimal size up to $O(1)$ factors. 
     Right panel: The NGS configurations in the $\{p(0),R\}$ plane, where $p(0)$ is the central pressure. $p(0)$ is plotted in units of the ($f$-independent) maximal pressure given in \Eq{eq::BM1_max_pressure}, while the radius is given in units of the ($f$-dependent) maximal radius expected for the SBOs, given in \Eq{eq::rmin_rmax_sbo_BM2}. The numerical results agree with the analytical estimates of the maximal pressure and size up to $O(1)$ factors.}
     \label{fig::SBO_plots}
\end{figure}
\\ \\
As shown in \Fig{fig::SBO_plots}, the numerical solutions agree with our simple modelling of the SBOs given above. The smallest SBOs are the densest and exhibit the highest internal pressure. These properties are in fact $f$-independent, \eg the maximal pressure is given in this case by 
\begin{align}
\label{eq::BM1_max_pressure}
p^{\text{\tiny SBO}}_{\text{\tiny max}} \simeq g m_N \rho_\ngs \approx \varepsilon_0/2 \,.
\end{align}
for the BM1 benchmark. The maximal density can be found by solving $\Delta p_{\text{\tiny grad.}} = p^{\text{\tiny SBO}}_{\text{\tiny max}}$ numerically, which results in $\rho_{\text{\tiny max}} \approx 10 \rho_\ngs$. 
These results match, up to $O(1)$ factors, the numerical results shown in \Fig{fig::SBO_plots}, confirming the $f$-independent behavior of the smallest SBOs in number density (left panel) and  pressure (right panel). 
Note that the smallest SBOs are the least compatible with the underlying assumptions of our modeling of SBOs, namely constant density and small transition region. 
\Fig{fig::SBO_plots} also confirms our $f$-dependent predictions, \ie the minimal and maximal size of the SBOs. The qualitative behavior of the curves follows the description given above; the smallest SBOs with $R \simeq R^{\text{\tiny SBO}}_{\text{\tiny min}}$ are the densest and have the highest pressures. 
As the number density decreases and $\rho$ approaches $\rho_\ngs$, the object becomes larger, more dilute and the internal pressure decreases. 
This continues until $R \simeq R^{\text{\tiny SBO}}_{\text{\tiny max}}$, where gravity becomes important and matter must be added inside in order to counter the increasing gravitational pressure. 
From this point on, the mass and radius increase as the central number density and pressure increase. 
For both $f=5\times 10^{16}\GeV$ and $f= 10^{16}\GeV$, we find that the maximal size of the gravitationally-bound stars coincides with $R^{\text{\tiny SBO}}_{\text{\tiny max}}$, consistent with the analytic estimates in \Eq{eq::RGBO_max_estimate}.


\section{Large nucleon mass reduction in $f(\phi)G_{\mu\nu}^2$ models}
\label{app::fGG}

Prompted by \Eq{eq:anomaly}, we present here a simple model that can lead to a large reduction of the nucleon mass, thus populating most of the parameter space of \Fig{fig::ALP_phase_space}, albeit at the cost of some tuning. Above the QCD scale, we consider the following Lagrangian
\begin{align}
\mathcal{L}_{\phi} = \frac{1}{2} (\partial_\mu \phi)^2
- g {f}(\phi) \frac{\beta(g_s)}{2g_s} G^{\mu\nu}G_{\mu\nu}
- \epsilon \frac{\beta(g_s)}{g_s} \frac{ M_{\text{\tiny UV}}^4 }{32\pi^2} g {f}(\phi)\,,
\label{eq::L_Gsquared}
\end{align}
where ${f}(\phi)$ is a dimensionless function of the scalar field $\phi$ such that ${f}(0)=0$, and $g$ is a dimensionless coefficient (introduced in analogy to the ALP $g$ factor in \Eq{eq::m_V_ALP}).
The last term in \Eq{eq::L_Gsquared} is the UV contribution to the potential, naturally expected from closing a gluon loop at leading order in $g f(\phi)$, where $M_{\text{\tiny UV}}$ is the cutoff of the scalar theory and we allowed for tuning in the UV by introducing the parameter $\epsilon < 1$. 
Note that \Eq{eq::L_Gsquared} could be generalized by adding interaction terms with the light quarks $\sum_{q=u,d,s}f_q(\phi)m_q\bar{q}q$. 
As seen for the QCD axion in \Sec{sec::QCDaxion},
these interactions lead to a comparatively smaller coupling to nucleons, and we disregard them, along with quark masses, in the following.

Below the QCD scale, we use the well-known fact that the divergence of the dilatation current can be used to obtain the matrix element \cite{Shifman:1978zn,Kaplan:2000hh,Damour:2010rp}
\begin{align} 
\langle N | \frac{\beta(g_s)}{2g_s} G^{\mu\nu}G_{\mu\nu} | N \rangle = m_0\,,
\end{align}
where $m_0 \approx 869.5 \MeV$~\cite{Hoferichter:2015hva} is the nucleon mass in the chiral limit.
This allows to match the theory in \Eq{eq::L_Gsquared} to
\begin{align}
\mathcal{L}_{\phi}^{\text{\tiny IR}} \supset  -m_N\bar{N}N \left[1 - \frac{g}{g_*} f(\phi)\right]
- \left( \epsilon \frac{\beta}{g_s} \frac{M_{\text{\tiny UV}}^4}{32\pi^2} + 
c \Lambda_{\text{\tiny QCD}}^4 \right)
g f(\phi)\,,
\label{eq::L_IR}
\end{align}
at leading order in $g f(\phi)$ and neglecting the difference in $\beta/g_s$ between the scale at which \Eq{eq::L_Gsquared} is defined and the QCD scale, $\Lambda_{\text{\tiny QCD}}$.
Note that $g_* = m_N/m_0 \approx 1.08$, yet since we are neglecting quark masses we will consistently take $g_* = 1$.
We also added an IR contribution to the scalar potential in \Eq{eq::L_IR}, generated by the interaction of $\phi$ with the gluons and proportional to QCD contribution to the cosmological constant, which we have estimated as
\begin{align}
\langle 0 | \frac{\beta(g_s)}{2g_s} G^{\mu\nu}G_{\mu\nu} | 0 \rangle = c \Lambda_{\text{\tiny QCD}}^4\,,
\end{align}
with $c = O(1)$.
\\ \\
At this point, we can map the function $f(\phi)$ to either the ALP model of \Sec{sec::ALP_model}, with $f(\phi) = (1-\cos \phi/f)/2$, or to the linearly and quadratically coupled models of \Secs{sec::linear_coup_model}{sec::quad_coupling}, with $g f(\phi) = \phi/\mphi{}$ and $g f(\phi) = (\phi/\mphi{})^2$, respectively.
Let us note that in deriving \Eq{eq::L_IR} we kept only the leading order term in $g f(\phi)$, while a large (in-medium) reduction of the nucleon mass requires $g f(\phi) \sim 1$.
As a result, for small values of $m_*(\theta_\infty)$ (recall $\theta_\infty$ is the high-density limiting value of $\phi/f$), 
control over such non-linear terms is required.

Finally, evaluating the scalar potential in \Eq{eq::L_IR} at $\theta_\infty$, we find
\begin{align}
V(\theta_{\infty}) = \left( \epsilon \frac{\beta}{g_s} \frac{M_{\text{\tiny UV}}^4}{32\pi^2} + c \Lambda_{\text{\tiny QCD}}^4 \right) \left(1-\frac{m_*(\theta_{\infty})}{m_N}\right) 
\equiv V_0(\theta_{\infty}) \left(1-\frac{m_*(\theta_{\infty})}{m_N}\right)\,.
\label{eq::V_m_Gsquared}
\end{align}
By canceling the UV and QCD contributions to the potential, a significant fraction of the parameter space of \Fig{fig::ALP_phase_space} is populated.
In particular, note that the larger the degree of tuning, the smaller $V(\theta_{\infty})$ can be, which is the regime where larger departures (\eg in the mass and radius of NSs) are found w.r.t.~the standard GR prediction.
In \Fig{fig::ALP_phase_space}, the blue curve correspond to \Eq{eq::V_m_Gsquared} with $V_0(\theta_{\infty})\sim (0.17 \GeV)^4$.


\bibliography{bib/EOS}
\bibliographystyle{jhep}

\end{document}